\documentclass[
  10pt,
  twocolumn,
  reprint,
  superscriptaddress,
  nofootinbib,
  amsmath,
  amssymb,
  aps,
  prc,
  longbibliography,
  numerical,
  floatfix
]{revtex4-2}

\usepackage{graphicx}
\usepackage{dcolumn}
\usepackage{bm}
\usepackage{siunitx}
\usepackage[colorlinks=true, allcolors=blue]{hyperref}
\newcommand{\nicefrac}[2]{#1/#2}
\usepackage{orcidlink}
\usepackage{multirow}
\usepackage[normalem]{ulem}
\usepackage{dcolumn}
\allowdisplaybreaks

\newcommand{\eg}{\textit{e.g.}}

\newcommand{\etal}{\textit{et al.}}

\begin{document}

\preprint{APS/123-QED}

\title[Reevaluation of the Nuclear Charge Radii of Sr Isotopes]{Nuclear Charge Radii of Sr Isotopes: Reevaluation based on Transition Frequency Measurements in the $5s-5p-4d$ manifold in Sr$^+$ 
}

\author{J. Palmes\orcidlink{0009-0007-3928-7095}}
    \email{jpalmes@ikp.tu-darmstadt.de}
    \affiliation{Institut für Kernphysik, Technische Universität Darmstadt, Darmstadt, Germany}
\author{K. König\orcidlink{0000-0001-9415-3208}}
    \affiliation{Institut für Kernphysik, Technische Universität Darmstadt, Darmstadt, Germany}
    \affiliation{Helmholtz Forschungsakademie Hessen für FAIR, GSI Helmholtzzentrum für Schwerionenforschung, Darmstadt, Germany}
\author{B. K. Sahoo \orcidlink{0000-0003-4397-7965}}
 \affiliation{Atomic, Molecular and Optical Physics Division, Physical Research Laboratory, Navrangpura, Ahmedabad 380058, Gujarat, India}
\author{H. Bodnar\orcidlink{0009-0005-3056-3124}}
    \affiliation{Institut für Kernphysik, Technische Universität Darmstadt, Darmstadt, Germany}
\author{A. Candiello\orcidlink{0000-0002-3473-4745}}
    \affiliation{Instituut voor Kern- en Stralingsfysica, KU Leuven, Leuven, Belgium}
\author{A. Dorne\orcidlink{0000-0002-4014-5816}}
    \altaffiliation{Current address: Department of Physics, University of Strathclyde, Glasgow, United Kingdom}
    \affiliation{Instituut voor Kern- en Stralingsfysica, KU Leuven, Leuven, Belgium}
\author{R. de Groote\orcidlink{0000-0003-4942-1220}}
    \affiliation{Instituut voor Kern- en Stralingsfysica, KU Leuven, Leuven, Belgium}
\author{P. Imgram\orcidlink{0000-0002-3559-7092}}
    \altaffiliation{Current address: Helmholtz Forschungsakademie Hessen für FAIR, GSI Helmholtzzentrum für Schwerionenforschung, Darmstadt, Germany}
    \affiliation{Institut für Kernphysik, Technische Universität Darmstadt, Darmstadt, Germany}
    \affiliation{Instituut voor Kern- en Stralingsfysica, KU Leuven, Leuven, Belgium}
\author{I. Lopp\orcidlink{0009-0005-2643-8662}}
    \affiliation{Institut für Kernphysik, Technische Universität Darmstadt, Darmstadt, Germany}
\author{P. Müller\orcidlink{0000-0002-4050-1366}}
    \affiliation{Institut für Kernphysik, Technische Universität Darmstadt, Darmstadt, Germany}
\author{W. Nörtershäuser \orcidlink{0000-0001-7432-3687}}
    \affiliation{Institut für Kernphysik, Technische Universität Darmstadt, Darmstadt, Germany}
    \affiliation{Helmholtz Forschungsakademie Hessen für FAIR, GSI Helmholtzzentrum für Schwerionenforschung, Darmstadt, Germany}
\author{B. Ohayon\orcidlink{0000-0003-0045-5534}}
    \affiliation{Technion – Israel Institute of Technology, Haifa, Israel}
\author{R. Van Duyse\orcidlink{0009-0003-6034-2184}}
    \affiliation{Instituut voor Kern- en Stralingsfysica, KU Leuven, Leuven, Belgium}

\begin{abstract}
High-precision quasi-simultaneous collinear/anticollinear laser spectroscopy was performed to measure the $5s$ $^2S_{\nicefrac{1}{2}}\rightarrow 5p$ $^2P_{\nicefrac{1}{2}}$ (D1), the $5s$ $^2S_{\nicefrac{1}{2}}\rightarrow 5p$ $^2P_{\nicefrac{3}{2}}$ (D2), and the three $4d\rightarrow 5p$ transitions in naturally abundant Sr$^+$ isotopes. For absolute transition frequencies, an uncertainty as low as \SI{600}{\kilo\hertz} was achieved, while common-mode rejection allowed us to extract isotope shifts with uncertainties down to a level of \SI{200}{\kilo\hertz}, one order of magnitude better than previously achieved. Similarly, the uncertainties of the hyperfine-structure coefficients for $^{87}$Sr of the $5p$ states and the $4d$ $^2D_{\nicefrac{3}{2}}$ levels are improved. A King plot analysis yielded a field-shift ratio of the D2 and D1 lines of $F_\text{D2}/F_\text{D1}=1.004(5)$, which lies within the theoretically allowed region and can be used as a benchmark for atomic structure theory calculations. We use the information from all stable isotopes in the investigated transitions to compare field-shift and mass-shift constants obtained by various techniques regularly used in the literature, ranging from King-plots with purely experimental input to ab initio atomic structure calculations by state-of-the-art theory. We show that in the region above $N=50$, the charge radii are strongly dependent on the approach being used.
\end{abstract}

\maketitle

\section{\label{sec:Introduction}Introduction}
Alkaline atoms and singly charged alkaline-earth ions\footnote{The term `ion' corresponds in the following always to the singly charged positive ionic state if not explicitly stated otherwise.} have a single valence electron and no open subshells. Therefore, they have long been used as benchmarks for atomic structure calculations \cite{condon1935theory}. Transition frequencies, isotope shifts (IS), and hyperfine-structure (HFS) splittings are important observables of these systems since they can be measured with high precision and are used to extract, \eg, information about nuclear structure \cite{Campbell.2016,Yang.2023}. Isotope shifts have provided  nuclear charge radii for stable and short-lived isotopes of all alkaline \cite{Ewald.2004,sanchez_nuclear_2006,Huber.1978,Bonn.1979,Kreim.2014,Koszorus.2021,Thibault.1981,Schinzler.1978, Mane.2011c, Coc.1985,Collister.2014, Lynch.2014} and alkaline-earth elements \cite{Nortershauser.2009,Krieger.2012,Yordanov.2012, MartenssonPendrill.1992,GarciaRuiz.2016,Miller.2019,Mueller.1983, Buchinger.1985, Ahmad.1983,Ahmad.1988}, HFS splittings contain information on the nuclear spin \cite{Arnold.1987,Neyens.2005,Papuga.2013}, the nuclear magnetic dipole moment \cite{Arnold.1987,Geithner.1999,Takamine.2014,Yordanov.2007}, and the nuclear electric quadrupole moments \cite{Arnold.1992}. 

Extracting nuclear observables from isotope shifts and hyperfine splittings requires electronic factors obtained from atomic-structure calculations. The reliability of the resulting nuclear charge radii and electromagnetic moments therefore depends directly on the accuracy of these calculations.

Singly charged alkaline-earth elements provide an opportunity to test the calculated field- and mass-shift factors experimentally. Mg, Ca, Sr, and Ba each have at least three stable isotopes for which independent nuclear charge-radius information is available from muonic-atom spectroscopy and elastic electron scattering. These data enable a King-plot determination of the mass- and field-shift factors, which can be compared directly with atomic-structure calculations \cite{King.1984}. For alkaline elements, by contrast, the small number of stable isotopes generally precludes such a determination, and the extraction of their nuclear charge radii relies more strongly on calculated atomic factors \cite{Katyal2025}. Their isoelectronic alkaline-earth ions therefore serve as valuable benchmark systems for assessing these calculations.

Reliable atomic factors and well-founded uncertainties are important not only for the extraction of nuclear charge radii, which is receiving increasing attention across several fields \cite{Angeli.2026}, but also for precision tests of the Standard Model. In particular, the interpretation of atomic parity-nonconservation measurements in Cs in terms of the weak nuclear charge and the Weinberg angle is limited by the accuracy of the underlying atomic-structure calculations \cite{Bouchiat.1985,Bouchiat.1986,Wood.1997,Bennett.1999, Bouchiat.1997,Dzuba.2012,Toh.2019,Tah.2023,Sanamyan.2023}.

Ca$^+$ ions served as a first system to search for new bosons \cite{Berengut.2018} that mediate an interaction between electrons and neutrons beyond standard model (BSM) physics -- a field that has recently gained much interest in theory and experiment; see \eg~\cite{Door.2025}. Similar theoretical and experimental investigations are now proposed for transitions in Sr$^+$ ions \cite{Munro-Laylim.2022} and have already been carried out in Sr atoms \cite{Miyake.2019}. Furthermore, transitions in Sr$^+$ ions serve as atomic-clock candidates \cite{Steinel.2023} and for quantum computing \cite{Manovitz.2022}. Moreover, spectroscopy on Sr$^+$ in an ion trap has recently been proposed for environmental studies to detect the radiotoxic isotope $^{90}$Sr \cite{Jung.2017,Jung.2017b} and to surpass previously used optical detection techniques \cite{Wendt.1997,Bushaw.2000,Lu.2003}. The transition frequencies, isotope shifts, and hyperfine structure measurements reported here are also of importance for these endeavors.

Furthermore, the influence of the relativistic $S_{\nicefrac{1}{2}}$ contribution to the small component of the $P_{\nicefrac{1}{2}}$ wavefunction has previously been identified through a deviation in the field-shift ratio of the D1 and D2 transitions\footnote{It is common to refer to $nS_{\nicefrac{1}{2}} \rightarrow nP_{\nicefrac{1}{2}}$ and $nS_{\nicefrac{1}{2}} \rightarrow nP_{\nicefrac{3}{2}}$ transitions in alkaline-like systems as D1 and D2 lines due to their similarity to the corresponding transitions in sodium.} 
in the spectroscopy of Ba$^+$ ions \cite{Wendt.1988, Imgram.2019}. A corresponding high-precision measurement in Ca$^+$ led to a surprisingly large ratio that could not be explained by theory \cite{Shi.2016} but was later clarified in a new measurement that was in agreement with theory and consistent with measurements of the isotope shift in the $4s \rightarrow 3d$ transitions \cite{Muller.2020}.
Our measurements provide the first determination of the size of this effect in Sr. 

The structure of this paper is as follows: After a short description of the experimental setup and method in Sec.\,\ref{sec:Method}, we present the measurements and analysis of transition frequencies, IS and HFS in the $5s\,^2\!S_{\nicefrac{1}{2}}\rightarrow 5p\,^{2\!}P_{\nicefrac{1}{2},\nicefrac{3}{2}}$ (D1, D2)-- including the field-shift ratio -- and the $4d\,^{2\!}D_{\nicefrac{3}{2},\nicefrac{5}{2}}\rightarrow 5p\,^{2\!}P_{\nicefrac{1}{2},\nicefrac{3}{2}}$ transitions in Sr$^+$ ions in Secs.\,\ref{sec:Experiment}, \ref{sec:Results}. The hyperfine-structure splittings provide a benchmark test of state-of-the-art relativistic coupled-cluster calculations. Section \ref{sec:LeadOrdISParameters} is devoted to the determination of the isotope-shift parameters, which are then used in Sec.~\ref{sec:ChargeRadii} to extract a consistent set of charge radii for the stable and short-lived isotopes. 

\section{\label{sec:Method}Method}
The isotope shift
\begin{eqnarray}\label{eq:isotope_shift}
    \delta\nu_i^{A,A'} &\equiv& \nu_i^{A}-\nu_i^{A'} \\
    &\approx& 
    \frac{K_i}{\mu^{A,A'}} + F_i\delta\langle r_\mathrm{c}^2\rangle^{A,A'}, \text {with} \nonumber\\
    \mu^{A,A'} &=& \frac{M_A M_{A'}}{M_A-M_{A'}}
\end{eqnarray}
between two isotopes $A$ and $A'$ arises from the change in the atomic masses $M$ and the change in the mean-square nuclear charge radius $\delta\langle r_\mathrm{c}^2\rangle^{A,A'}$. Here, $K_i$ and $F_i=-Ze^2\Delta|\Psi(0)|^2_i/(6h\epsilon_0)$ denote the mass-shift and field-shift factors of transition $i$, with $\Delta|\Psi(0)|^2_i$ the transition-specific change in the electron density at the nucleus.

After multiplication with the modified mass factor  $\mu^{A,A'}$ 
one obtains a linear relation between the mass-modified isotope shift and the mass-modified change in charge radius,
\begin{equation}\label{eq:KingPlot2}
    \mu^{A,A'}\delta\nu_i^{A,A'} \approx K_i + F_i\,\mu^{A,A'}\delta\langle r_\mathrm{c}^2\rangle^{A,A'}.
\end{equation}
If $\delta\langle r_\mathrm{c}^2\rangle^{A,A'}$ is known from independent methods, this representation allows for a direct determination of the atomic factors $K_i$ and $F_i$ (see Sec.~\ref{sec:LeadOrdISParameters}). Conversely, once these factors are known, Eq.~\eqref{eq:KingPlot2} can be used to improve the charge-radius information of isotopes with known radii or to extract the charge radii of other, in particular, short-lived isotopes, as described in detail in Ref.~\cite{Fricke.2004} and applied here in Sec.~\ref{sec:ChargeRadii}.

Alternatively, by measuring the isotope shifts for two transitions $i$ and $j$, the change in the mean-square nuclear charge radius in Eq.~\eqref{eq:isotope_shift} can be eliminated yielding
\begin{equation}\label{eq:FSR}
    \mu^{A,A'}\delta\nu_i^{A,A'} \approx K_i - \frac{F_i}{F_j}K_j + \frac{F_i}{F_j}\mu^{A,A'}\delta\nu_j^{A,A'}.
\end{equation}
This is a linear relation between the measured mass-modified isotope shifts, with a slope given by the ratio of the field-shift factors, $F_i/F_j$, of the two transitions, the so-called King plot. Since this ratio can be accessed through calculations of the electronic wavefunctions at the nucleus for all involved states, it provides a benchmark for comparing atomic-structure theory with experiment \cite{Wendt.1984,Shi.2016,Imgram.2019,Muller.2020}. This is particularly feasible in systems with a single valence electron outside closed shells, such as Sr$^+$, since such systems are well suited for coupled-cluster calculations, which are among the most accurate methods in atomic-structure theory.

\begin{figure}
    \centering
    \includegraphics[width=\linewidth]{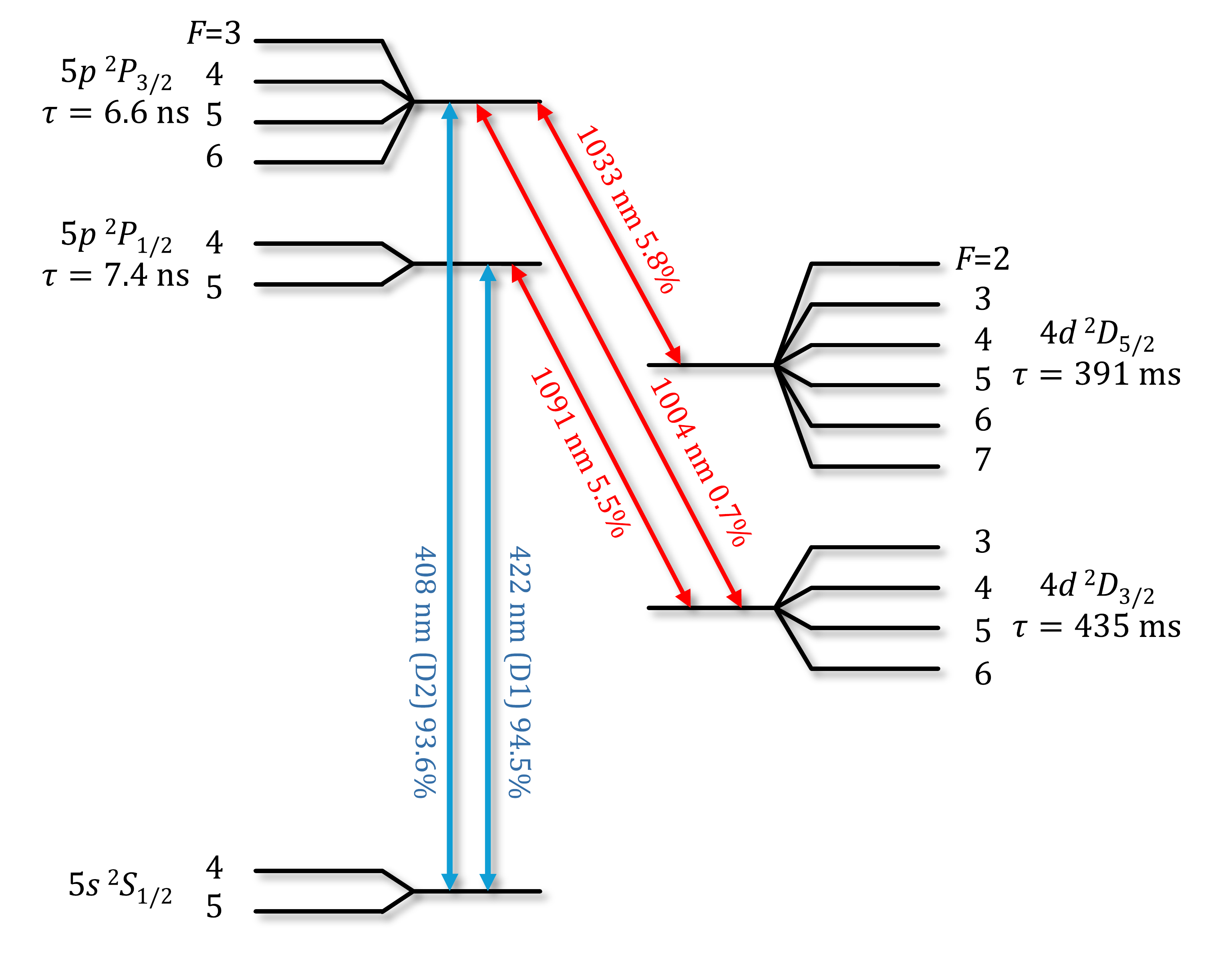}
    \caption{Level scheme and transitions of Sr$^+$ ions relevant for this work. The hyperfine substates of $^{87}$Sr are indicated.}
    \label{fig:level_scheme}
\end{figure}

Strontium has four naturally abundant isotopes, namely $^{88-86}$Sr and $^{84}$Sr. Challenges for laser spectroscopy are the comparatively low abundance of 0.56(2)\,\% in the case of $^{84}$Sr and the nuclear spin of $I=\nicefrac{9}{2}$ for $^{87}$Sr, which causes hyperfine splitting with only partially resolved lines in some of the transitions.
The ground state of Sr$^+$ is the $5s\,^2\!S_{\nicefrac{1}{2}}$ level with a single valence electron outside of the closed krypton core. Two dipole transitions at 408\,nm (D1) and 422\,nm (D2) lead to the $5p\,^2\!P_{\nicefrac{1}{2}}$ (23715.19\,cm$^{-1}$) and $5p\,^2\!P_{\nicefrac{3}{2}}$ (24516.65\,cm$^{-1}$) levels with lifetimes of a few ns \cite{Moore1971}. Both levels decay predominantly back to the ground state but have weak branches into the metastable $4d$ $^2\!D_{\nicefrac{3}{2},\nicefrac{5}{2}}$ levels (14555.90\,cm$^{-1}$ and 14836.24\,cm$^{-1}$) \cite{Moore1971}. Once populated, the electron in the $4d$-levels can be excited by one of three transitions at 1004\,nm, 1032\,nm or 1091\,nm into the $5p$ manifold as shown in Fig.\,\ref{fig:level_scheme}. 

In this work, collinear laser spectroscopy (CLS) was chosen  as it can determine isotope shifts in allowed dipole transitions to better than 1\% of their natural linewidth \cite{Imgram.2019, Muller.2020}. 
In CLS, an ion beam with energy of a few 10\,keV is superimposed either collinearly (c) or anticollinearly (a) with the laser. The electrostatic acceleration out of an ion source leads to a strong compression of the thermal velocity distribution in the direction of flight and therefore strongly reduces the Doppler broadening \cite{Kaufman.1976}. However, the collinear geometry also leads to maximal Doppler shifts
\begin{equation}
    \nu_\text{c/a} = \nu_0 \gamma \left(1\pm\beta\right),
\end{equation}
where $\nu_0$ is the transition frequency in the rest-frame of the ion, $\nu_\text{c/a}$ is the laser frequency in the laboratory-frame for collinear or anticollinear geometry, $\gamma=1/\sqrt{1-\beta^2}$ and $\beta=v/c$ is the velocity of the ions in the lab-frame divided by the speed of light. Instead of scanning the laser frequency, the ion velocity can be scanned, which varies the effective Doppler shift, to obtain spectra by scanning an additional applied voltage $U_\text{scan}$ to the interaction region. Another advantage of the collinear technique in the case of the Sr isotopes is a clear separation of isotope-specific peaks in the spectra due to the mass and, hence, velocity differences, while the natural isotope shifts are very small due to the large cancellation of field-shift and mass-shift contributions in this mass region.

Absolute transition frequencies can be determined velocity-independently by performing measurements in collinear and anticollinear direction \mbox{(quasi-)simultaneously}
\begin{equation}
    \nu_0=\sqrt{\nu_\text{c}\nu_\text{a}}=\sqrt{\nu_0\gamma\left(1+\beta\right)\nu_0\gamma\left(1-\beta\right)}.
\end{equation}
Performing those measurements in alternating order, namely anticollinear-collinear-collinear-anticollinear, cancels contributions from a drifting acceleration potential and will be further referenced as acca measurements.

\section{\label{sec:Experiment}Experiment}
\begin{figure}
    \includegraphics[width=\linewidth]{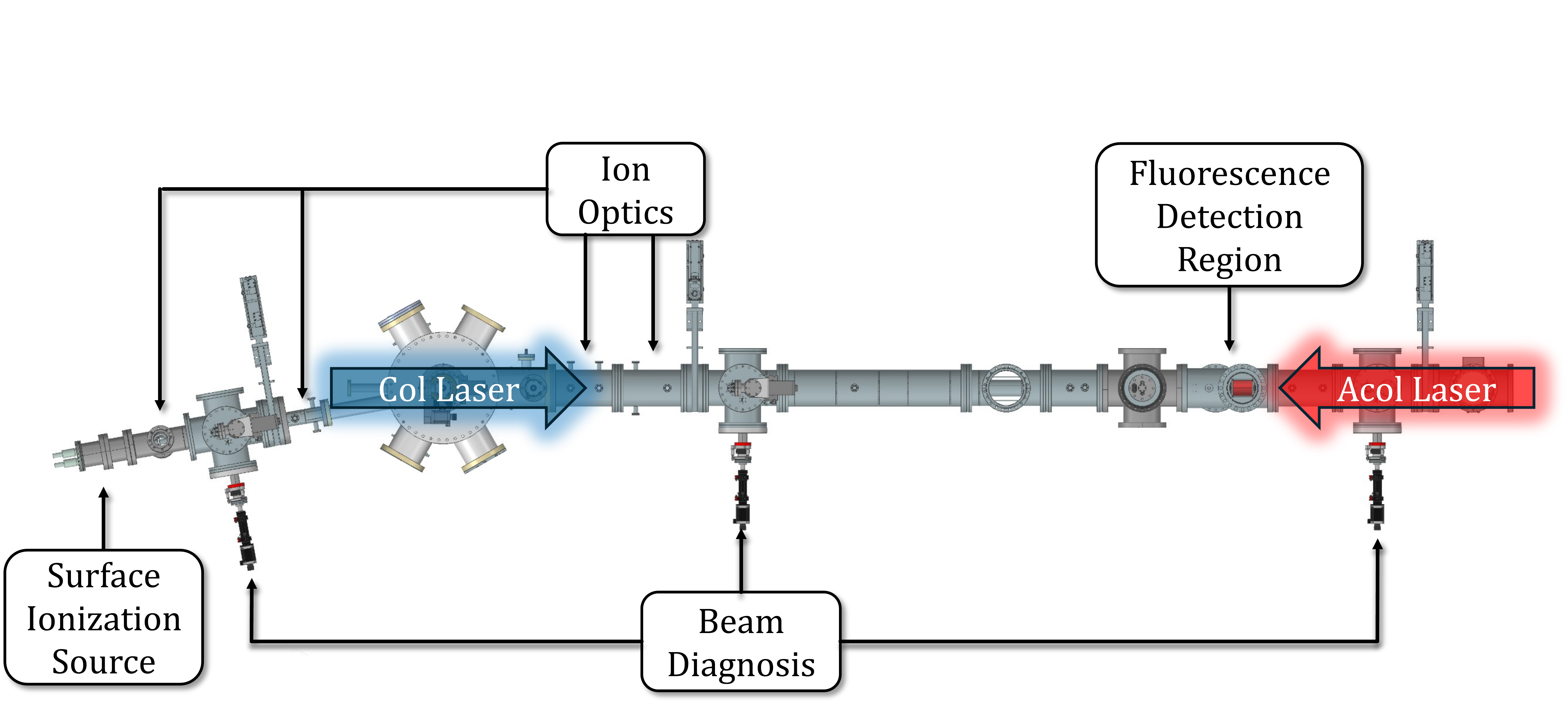}
    \caption{A schematic of the updated COALA beamline at TU Darmstadt. In this experiment, the Sr$^+$ ions are created in the surface ionization source. Ion optics and three beam diagnostic stations equipped with Faraday cups and multi-channel plates allow superimposing the ion and laser beams in the fluorescence detection region.}
    \label{fig:setup}
\end{figure}
The experiment was performed at the Collinear Apparatus for Laser Spectroscopy and Applied Science (COALA) shown in Fig.\,\ref{fig:setup}. An extensive description of the COALA beamline can be found in \cite{Konig.2020b} with a recent update provided in \cite{Imgram.2023c}. Similar measurements were previously reported on Ca$^+$ \cite{Muller.2020} and Ba$^+$ \cite{Imgram.2019}. Three main components need to be precisely controlled to reach an accuracy of the order of 100\,kHz: the ion velocity, the laser frequency, and the overlap of the laser beams. 
The velocity of the ions inside the fluorescence detection region (FDR) only depends on the initial starting potential $U_\text{start}$ and the scanning voltage $U_\text{scan}$ applied to the FDR. The starting potential of approximately 20\,kV was measured via a custom high-voltage divider \cite{Konig.2024} with a divider ratio of about 2000:1 and variations of $\leq 0.2$\,ppm on the time scale of a measurement. The voltage of about 10\,V across the precision resistor of the chain was measured with a 8.5 digit multimeter (Keysight 3458A), which controls a feedback loop consisting of a custom digital-to-analog converter to stabilize the starting potential, improving the stability of the beam velocity by more than one order of magnitude \cite{Konig.2024}.
The scanning voltage was provided by a $\pm 10$\,V 18-bit digital-to-analog converter (Analog Devices, AD5781) which was amplified by a factor of 50 with a Kepco BOP500M linear bipolar amplifier. Recently, we have quantified the influence of field penetrations into the FDR \cite{Konig.2024}. In acca measurements of the even-even isotopes, this has been circumvented by choosing both laser frequencies such that the resonances occur at the same scanning voltage, ideally at very low voltages, as the field penetration effects scale proportional to the applied voltage. For wider spectra, such as the hyperfine structure of $^{87}$Sr, this effect resulted in a relative uncertainty of the peak spacings of about 40 ppm.

Laser light for the $4d \,^{2\!}D_{\nicefrac{3}{2},\nicefrac{5}{2}}\rightarrow 5p\,^{2\!}P_{\nicefrac{3}{2}}$ 
transitions (1004 nm and 1033 nm) was produced by two Matisse 2 TS Ti:Sapphire ring lasers, each pumped by a Millennia eV diode-pumped Nd:YVO$_4$ laser. For the D1 (422 nm) and D2 (408 nm) transitions, the output of the Ti:Sapphire lasers was frequency doubled with a WaveTrain 2 using lithium triborate (LBO) crystals. Both Matisse lasers were frequency stabilized using an FC1500-250-WG frequency comb with a GPS-disciplined quartz oscillator and frequency drifts of the lasers were compensated by adjusting their reference-cavity lengths.
This reduces the frequency uncertainty and drifts typically down to the level of 30 kHz \cite{Konig.2020b}, limited by the short-term stability of the Matisse.

For the $4d$ $^2\!D_{\nicefrac{3}{2}}\rightarrow 5p$ $^2\!P_{\nicefrac{1}{2}}$ transition at 1091 nm a tuneable diode laser of type Toptica DL Pro was stabilized to a HighFinesse WSU30 wavemeter. The wavemeter was calibrated with a stable He:Ne laser, allowing an absolute frequency determination with a 1$\sigma$-uncertainty of 10\,MHz.

The laser light was transported to the experiment via single-mode optical fibers, providing collimated beams with Gaussian profiles and FWHM diameters of approximately 1\,mm, while reducing the vibrations observed with free-space beam transport. Laser powers of approximately 5\,$\mu$W and 600\,$\mu$W were used for the $5s\rightarrow5p$ and $4d\rightarrow5p$ transitions, respectively, corresponding to peak intensities of approximately 0.44 and 53\,mW/cm$^2$. Based on the natural linewidths and branching fractions \cite{zhang_iterative_2016}, the estimated saturation intensities are approximately 38 and 47\,mW/cm$^2$ for the D1 and D2 transitions, respectively, and approximately 37, 490, and 51\,mW/cm$^2$ for the $D_{\nicefrac{3}{2}}\rightarrow P_{\nicefrac{1}{2}}$, $D_{\nicefrac{3}{2}}\rightarrow P_{\nicefrac{3}{2}}$, and $D_{\nicefrac{5}{2}}\rightarrow P_{\nicefrac{3}{2}}$ transitions. These estimates neglect transition-specific polarization and magnetic-sublevel factors. For the D1 and D2 transitions, measurements were performed at several laser powers, and the operating power was chosen to provide sufficient signal-to-noise ratio while minimizing the uncertainty of the fitted line position. The selected intensity was less than 2\% of the estimated saturation intensities, such that power broadening is expected to be negligible. By contrast, the intensity used for the infrared transitions is comparable to the estimated saturation intensities of the $D_{\nicefrac{3}{2}}\rightarrow P_{\nicefrac{1}{2}}$ and $D_{\nicefrac{5}{2}}\rightarrow P_{\nicefrac{3}{2}}$ transitions and may therefore result in noticeable power broadening. The higher infrared laser power was required to compensate for the small population of the metastable $4d$ states.

The overlap of the two laser beams was optimized by coupling both beams through the beamline into the fiber of the counter-propagating laser and maximizing the transmitted power. Given the 6.5\,m distance between the two fiber couplers and beam-spot diameters of approximately 2\,mm, a maximum angular deviation of 0.15\,mrad is estimated. Its effect on the measured absolute transition frequency depends on the transition wavelength but is typically of the order of 100\,kHz. The overlap of the ion beam with the laser beams was ensured using iris apertures distributed along the beamline, through which both the transmitted laser power and ion current were maximized, limiting angular deviations to approximately 1.2\,mrad. With the optimized counter-propagating geometry of the two laser beams, the corresponding effect is negligible, at approximately 10\,kHz.

Similar as for the $3d$ levels of Ca$^+$ ions in previous work \cite{Nortershauser.1998b}, we found the metastable $4d$ levels sufficiently populated in the surface ionization source. This is most probably caused by inelastic collisions with atoms during the acceleration out of the ion source, since the population of the $d$-states was further enhanced with increasing background pressure once the turbo-molecular pump nearest to the ion source was turned off. The signal was sufficient to record resonances of the $4d\rightarrow 5p$ transitions on a timescale of minutes.  
The detection of the blue photons emitted in the dominant $5p \rightarrow 5s$ decay following the $4d\rightarrow 5p$ excitation is almost background free since the photomultiplier tubes are not sensitive to the infrared laser light.

\section{\label{sec:Results}Results}
\subsection{Experimental spectra of Sr$^+$ transitions} 
\begin{figure*}[t]
    \centering
    \includegraphics[width=1\linewidth]{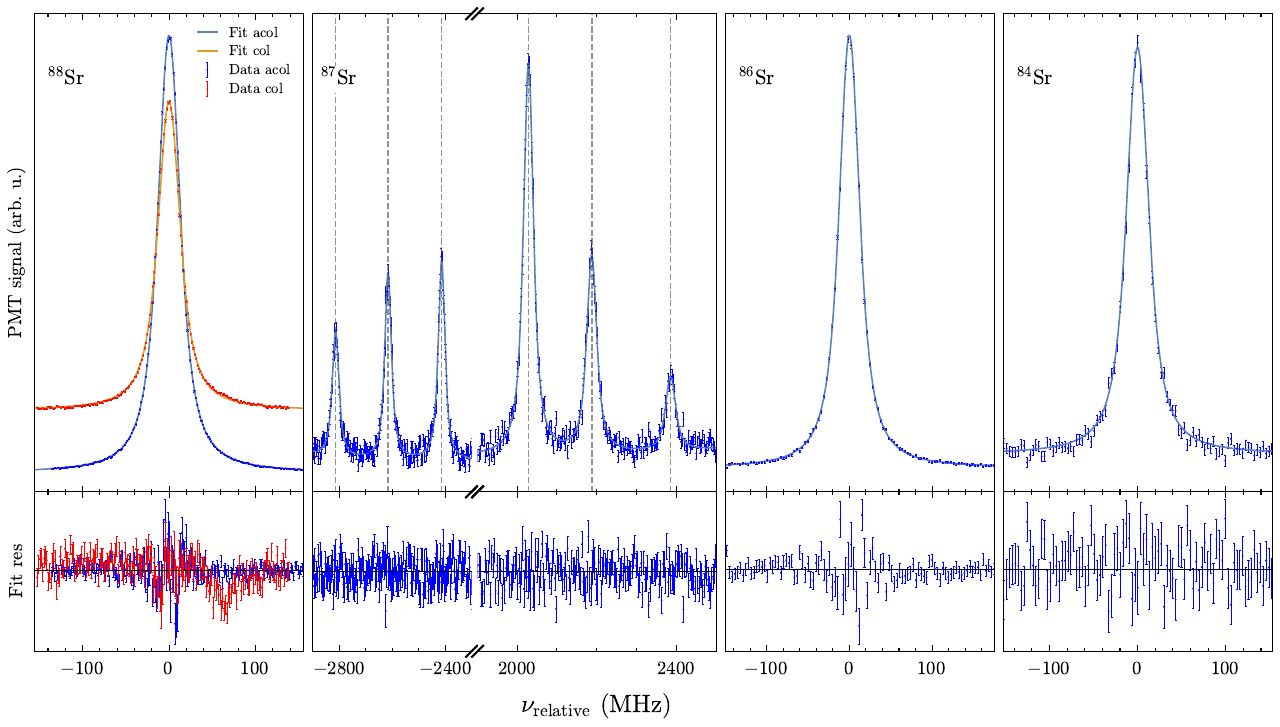}
    \caption{Experimental spectra of the $5s\,^2\!S_{\nicefrac{1}{2}}\rightarrow 5p\,^2\!P_{\nicefrac{3}{2}}$ transition (D2) in $^{88-86}$Sr and $^{84}$Sr fitted with Voigt-profiles to extract the transition frequency, the isotope shifts and the hyperfine structure $A$ and $B$ parameters of the upper and lower states of the transition. For $^{88}$Sr the spectra obtained in collinear (red) and anticollinear geometry (blue) are shown, whereas in the other cases only an anticollinear spectrum is depicted. For $^{87}$Sr, the position of each hyperfine peak is marked with a dashed line.}
    \label{fig:plot_d1}
\end{figure*}

Spectra of $^{88}$Sr, $^{87}$Sr, and $^{84}$Sr frequency measurements in the D2 transition are depicted in Fig.\,\ref{fig:plot_d1} as an example for all isotopes in the D1 and D2 transitions. For the most naturally abundant isotope $^{88}$Sr$^+$, spectra in collinear and anticollinear geometry are included in Fig. \ref{fig:plot_d1}, while the others are restricted to the anticollinear case for clarity reasons. The two triplets in the HFS of $^{87}$Sr$^+$  are separated by more than 4\,GHz due to the large hyperfine splitting in the $5s$ ground state. The spectrum of the least abundant isotope $^{84}$Sr (0.56\%) provides an impression of the data quality. All peaks were fitted with symmetric Voigt profiles. The Lorentzian component dominates, yielding linewidths of $\Gamma_\mathrm{D1}=21.5(3)$~MHz and $\Gamma_\mathrm{D2}=24.1(2)$~MHz, both close to the previously reported natural linewidths of $\Gamma_{\mathrm{nat,D1}}=20.4(2)$~MHz \cite{likforman_precision_2016} and $\Gamma_{\mathrm{nat,D2}}=22.7(3)$~MHz \cite{zhang_iterative_2016}. Given the low saturation parameters discussed above, the slight excess over the natural linewidths cannot be fully explained by power broadening. The Gaussian contribution of approximately $\sigma=5(1)$\,MHz for both transitions is comparatively small and represents the residual inhomogeneous broadening in the experiment. Nevertheless, small systematic structures remain visible in the fit residuals of $^{88}$Sr, suggesting that the measured line shape is not perfectly described by a symmetric Voigt profile. However, the influence of these small line-shape asymmetries largely cancels when combining the collinear and anticollinear measurements and therefore does not significantly affect the determined rest-frame transition frequency \cite{Krieger2017}.

To investigate this possibility, a more sophisticated line-shape model accounting for the thermal ion velocity distribution in the source \cite{Kretzschmar.2004, Muller.2020} was benchmarked on a test set comprising more than 200 spectra using the \texttt{SciPy} package \cite{virtanen2020scipy}. However, compared to the symmetric Voigt profile, this model increased the average reduced $\chi^2$ from 1.2 to 1.6 and was therefore not adopted for the final analysis.

Besides the acca measurements, in which the collinear and anticollinear resonances of one isotope were measured sequentially, we employed a second scheme referred to as ``direct IS''. Here, the laser was kept at a fixed frequency while the voltage applied to the fluorescence detection region was scanned over a range of up to 1\,kV. This allowed resonances of several isotopes to be recorded within the same voltage scan and their isotope shifts to be determined directly from the relative peak positions. In this approach, constant frequency offsets largely cancel in the extracted isotope shifts. The small residual dependence on the laser frequency, caused by the slightly different Doppler factors of the isotopes, was included in the uncertainty determination. Possible frequency changes during the scan and uncertainties of the applied scanning voltage were treated separately.

\begin{figure*}[t]
    \centering
    \includegraphics[width=1\linewidth]{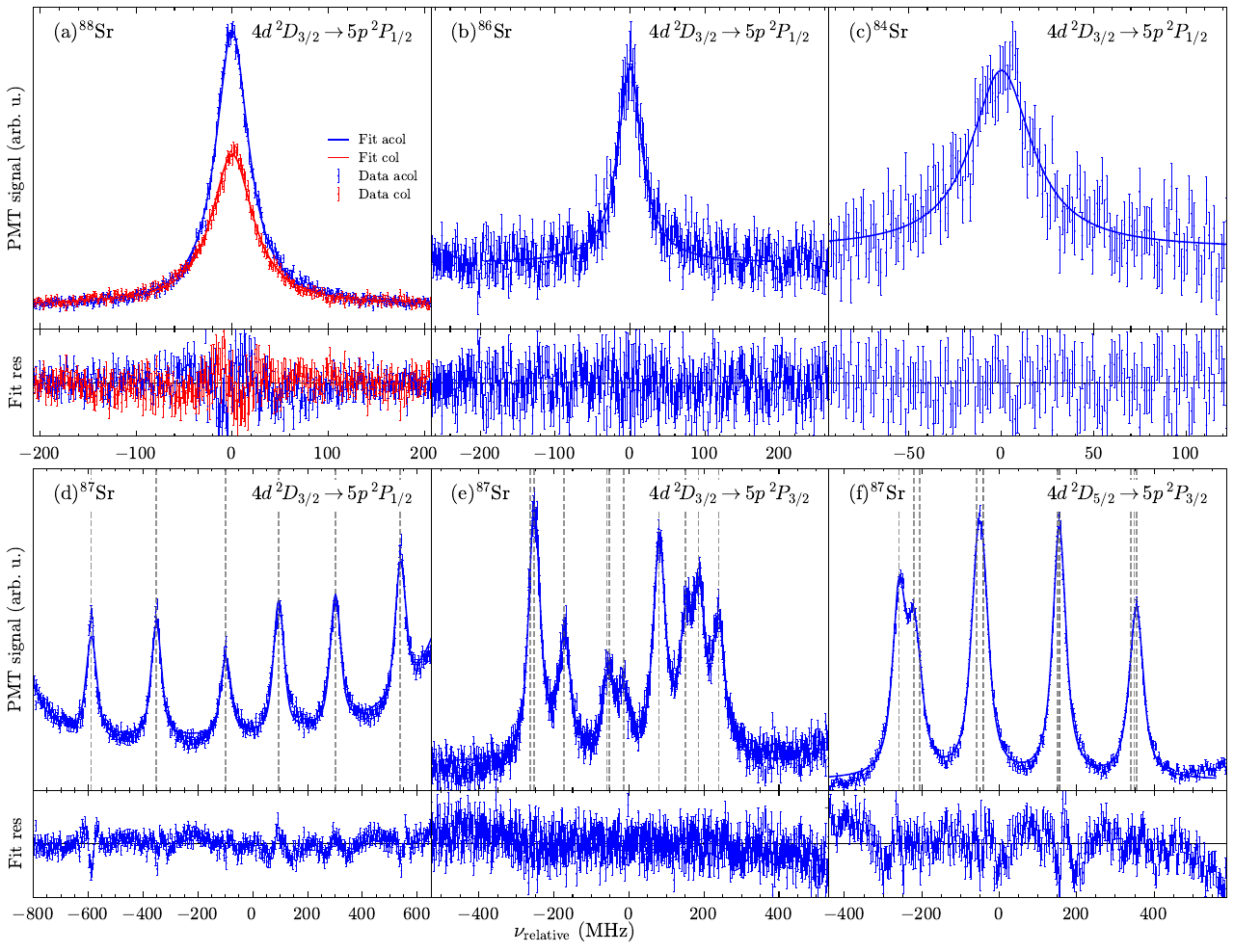}
    \caption{Experimental spectra of the $4d\,^2{\!}D\rightarrow 5p\,^{2\!}P$ transitions with symmetric Voigt-profile fits. As an example, a spectrum for both geometries for the $4d\,^2{\!}D_{\nicefrac{3}{2}}\rightarrow 5p\,^{2\!}P_{\nicefrac{3}{2}}$ transition is given for $^{88}$Sr in (a). In (b) and (c) just the anticollinear spectra are shown for $^{86}$Sr and $^{84}$Sr respectively. In (d-f), the $^{87}$Sr hyperfine spectra of the different transitions are shown, with each individual hyperfine peak position marked by a dashed line. While for the $4d\,^2{\!}D_{\nicefrac{3}{2}}\rightarrow 5p\,^{2\!}P_{\nicefrac{1}{2}}$ transition the six hyperfine peaks are well separated in (d), the ten peaks for the $4d\,^2{\!}D_{\nicefrac{3}{2}}\rightarrow 5p\,^{2\!}P_{\nicefrac{3}{2}}$ transition in (e) are more difficult to resolve. Owing to the small hyperfine splittings, the assignment of the twelve peaks of the $4d\,^2{\!}D_{\nicefrac{5}{2}}\rightarrow 5p\,^{2\!}P_{\nicefrac{3}{2}}$ transition in (f) was guided by the previously measured hyperfine constants of the $5s\,^2{\!}S_{\nicefrac{1}{2}}\rightarrow 4d\,^{2\!}D_{\nicefrac{5}{2}}$ transition \cite{barwood_observation_2003}. In those transitions of $^{87}$Sr, the background is distorted by the nearby peaks of $^{88}$Sr and $^{86}$Sr, which were only included in the scan and fit for the $4d\,^2{\!}D_{3/2}\rightarrow 5p\,^{2\!}P_{1/2}$ transition. For the remaining two transitions, the effect is visible in the residuals.}
    \label{fig:plot_d3p3}
\end{figure*}

For the infrared $4d\to5p$ transitions, representative spectra of all stable even isotopes in the $4D_{\nicefrac{3}{2}}\rightarrow 5P_{\nicefrac{3}{2}}$ transition are shown in Fig.\,\ref{fig:plot_d3p3} (a--c). The hyperfine structure of $^{87}$Sr is in some transitions only partially resolved and the corresponding spectra are depicted in Fig.\,\ref{fig:plot_d3p3} (d--f). To compensate for the lower statistics caused by the small population of the $d$ states, the ion-source temperature was increased, resulting in broader observed resonance profiles. Depending on the heating current applied to the source, the measured linewidths range from 30 to 50\,MHz, remaining on the order of the natural linewidths: $\Gamma_{\mathrm{nat}}(4D_{\nicefrac{5}{2}}\rightarrow5P_{\nicefrac{3}{2}})=25.2(4)$\,MHz \cite{pinnington_studies_1995,letchumanan_lifetime_2005}, $\Gamma_{\mathrm{nat}}(4D_{\nicefrac{3}{2}}\rightarrow5P_{\nicefrac{3}{2}})=25.2(4)$\,MHz \cite{biemont_lifetimes_2000,pinnington_studies_1995}, and $\Gamma_{\mathrm{nat}}(4D_{\nicefrac{3}{2}}\rightarrow5P_{\nicefrac{1}{2}})=20.2(3)$\,MHz \cite{biemont_lifetimes_2000,pinnington_studies_1995}. The dependence of the observed linewidth on the source-heating current supports the interpretation that the small excess linewidth observed in the D1 and D2 transitions also originates predominantly from the ion source. Although the fitted Gaussian contribution in those measurements is only approximately $\sigma=5(1)$\,MHz, an asymmetric source-induced velocity distribution is not fully represented by the symmetric Voigt model and may therefore also contribute to the fitted Lorentzian width.

\subsection{Combination  of results}
Measurements of the same observables (transition frequencies, isotope shifts and hyperfine parameters) performed on different days or with different measurement schemes were first averaged within the respective set of measurements and then compared, searching for indications of systematic differences. Therefore, the weighted average 
\begin{equation}
    X = \frac{\Sigma_i x_i  \sigma_i^{-2}}{\Sigma_i \sigma_i^{-2}},
\end{equation}
within each set was calculated. Its uncertainty was obtained as the maximum of the estimator for the standard error of the weighted average and the standard error of the mean for the respective number of measurements
\begin{equation}
    \Delta X = \max\left( \left(\Sigma_i \sigma_i^{-2}\right)^{-1/2}, \left(\Sigma_i (x_i-\bar{x})^2/N\right)/\sqrt{N}\right).
\end{equation}

We observed jumps of the measured absolute transition frequencies in the range of a few hundred kHz at different days, which were traced back to the frequency comb. At the time of these measurements, the frequency-comb's femtosecond laser amplifier was suffering from an aging pumping diode, resulting in low output power that led to limited beat signals with the spectroscopy laser. In our case, we found that these limited beat signals caused a systematically lower rate of about 100 kHz for the frequency counter that evaluated the beat frequency. This can lead to a higher or lower frequency readout depending on the relative frequency difference of both lasers. Once set up, the beat intensity stayed constant during the day, explaining the day to day jumps. To validate the frequency scale, the $4s\,^2S_{1/2}\rightarrow4p\,^2P_{1/2}$ (397\,nm) transition in Ca$^+$ was measured using the same surface ionization source. No additional offset beyond the observed jumps was found \cite{Muller.2020}. Since these jumps were observed only between measurements performed on different days, they affect the absolute transition frequencies but are irrelevant for isotope shifts obtained from transition-frequency differences measured on the same day. In this case, the common day-dependent frequency offset cancels in the difference. We report the measured transition frequencies in Tab.\,\ref{tab:abs} and add a conservative systematic contribution of 600\,kHz to the absolute frequency uncertainty, covering the full span of the observed frequencies.

No other systematic differences between the different sets were observed or exceeded $2\sigma$ of the combined uncertainties.

\subsection{Transition frequencies}
The transition frequencies obtained from our measurements are listed in Tab.\,\ref{tab:abs}. For the $D_{\nicefrac{3}{2}}\rightarrow P_{\nicefrac{3}{2}}$ and $D_{\nicefrac{5}{2}}\rightarrow P_{\nicefrac{3}{2}}$ transitions, frequency-comb-supported acca measurements were performed only for $^{88}$Sr and $^{86}$Sr. For the $D_{\nicefrac{3}{2}}\rightarrow P_{\nicefrac{3}{2}}$ transition, precise transition frequencies of $^{87}$Sr and $^{84}$Sr were determined by combining the directly measured isotope shifts with the frequency-comb-referenced rest-frame frequency of $^{88}$Sr. For the $D_{\nicefrac{5}{2}}\rightarrow P_{\nicefrac{3}{2}}$ transition, the isotope shifts were obtained from wavemeter-referenced measurements of all stable isotopes performed with a diode laser. These isotope shifts were then combined with the frequency-comb-referenced rest-frame frequency of $^{88}$Sr measured separately using the Matisse laser. In both cases, the transition frequency of isotope $A$ is given by
\begin{equation}
    \nu_i^A = \nu_i^{88} + \delta\nu_i^{A,88}.
    \label{eq:absolute_from_direct_is}
\end{equation}
For statistically independent quantities, the corresponding uncertainty is
\begin{equation}
    \sigma^2(\nu_i^A)
    =
    \sigma^2(\nu_i^{88})
    +
    \sigma^2(\delta\nu_i^{A,88}).
\end{equation}

For the $D_{\nicefrac{5}{2}}\rightarrow P_{\nicefrac{3}{2}}$ transition, combining the diode-laser isotope shifts with the Matisse-based frequency-comb measurement yields absolute transition frequencies with uncertainties of approximately 1--2\,MHz.
For the remaining $D_{\nicefrac{3}{2}}\rightarrow P_{\nicefrac{1}{2}}$ transition, the laser wavelength of 1091\,nm exceeds the spectral range of the frequency comb and from previous investigations with our HighFinesse WSU30 we also expect the deviations in this range to be larger than the specified 10\,MHz 1-$\sigma$ uncertainty \cite{konig_performance_2020}. Thus, we determined the transition frequencies via
\begin{equation}
    \nu^A_{D_{\nicefrac{3}{2}}\rightarrow P_{\nicefrac{1}{2}}} = \nu^A_{D1} - \left( \nu^A_{D2} - \nu^A_{D_{\nicefrac{3}{2}}\rightarrow P_{\nicefrac{3}{2}}}\right). \label{eq:ring_closure}
\end{equation}
The uncertainties of this method are approximately one order of magnitude smaller compared to the wavemeter-based acca transition frequency measurements.

From the measured frequencies, we can also obtain transition frequencies for the $S_{\nicefrac{1}{2}}\rightarrow D_{\nicefrac{3}{2}, \nicefrac{5}{2}}$ transitions in all isotopes. The results are listed in Tab.\,\ref{tab:absCalc} and are in excellent agreement with literature values \cite{margolis_absolute_2003,lybarger_precision_2011,Steinel.2023}. Precise frequencies in the $5S_{\nicefrac{1}{2}}\rightarrow 5D_{\nicefrac{3}{2}}$ transition have -- to our knowledge -- not been reported so far. 
 
\begin{table*}[t]
    \centering
    \caption{Transition frequencies of all investigated transitions for all stable isotopes in MHz. \label{tab:abs}}
    \begin{ruledtabular}
    \sisetup{table-format = 9.1(2){$^\dagger$}}
    \begin{tabular}{S[table-format=2.0]*5S@{}}
        {A} & {$\nu_{D1}^A$ } & {$\nu_{D2}^A$ } & {$\nu^A_{D_{\nicefrac{3}{2}}\rightarrow P_{\nicefrac{1}{2}}}$ } & {$\nu^A_{D_{\nicefrac{3}{2}}\rightarrow P_{\nicefrac{3}{2}}}$ } & {$\nu^A_{D_{\nicefrac{5}{2}}\rightarrow P_{\nicefrac{3}{2}}}$ }\\
        \hline
        \rule{0mm}{3.5mm}
        88 & 710962838.3 (6) & 734989824.5 (6) & 274589143.7(11)$^\dagger$ & 298616129.9 (6) & 290210780.4 (6) \\
        87 & 710962781.1 (6) & 734989768.4 (6) & 274589334.9(15)$^\dagger$ & 298616322.0 (13)$^\ddagger$& 290210970.0 (23)$^\ddagger$ \\
        86 & 710962666.6 (6) & 734989653.5 (6) & 274589543.8(10)$^\dagger$ & 298616530.7 (6) & 290211179.3 (6)\\
        84 & 710962462.8 (6) & 734989449.5 (6) & 274589968.8(15)$^\dagger$ & 298616955.4 (13)$^\ddagger$ & 290211603.0 (15)$^\ddagger$\\
    \end{tabular}
    \end{ruledtabular}
        {\raggedright
    $^\dagger$ Calculated according to Eq. \ref{eq:ring_closure}.\\
    $^\ddagger$ Calculated from frequency comb supported acca measurement for $^{88}$Sr plus isotope shift: $\nu^A_i=\nu^{88}_i+\delta\nu^{A, 88}_i$.
    \par}
\end{table*}

\begin{table}[t]
\caption{Extracted $5s\,^2\!S_{\nicefrac{1}{2}} \rightarrow 4d\,^2\!D_{\nicefrac{3}{2},\nicefrac{5}{2}}$ frequencies from the measured $S_J\rightarrow P_J$ and $D_J \rightarrow P_J$ transitions for all stable isotopes in MHz.\label{tab:absCalc} }
    \centering
    \begin{ruledtabular}
    \begin{tabular}{S[table-format=2.0]*2S[table-format=9.6(2)]*1S[table-format=4.0]}
        {A} & {$\nu^A_{S_{\nicefrac{1}{2}}\rightarrow D_{\nicefrac{3}{2}}}$} & {$\nu^A_{S_{\nicefrac{1}{2}}\rightarrow D_{\nicefrac{5}{2}}}$} & $\text{Ref.}$ \\
        \hline
        \rule{0mm}{3.5mm}
        88 & 436373694.6 (9)  & 444779044.0	(8) &\\
           &                  & 444779044.095485 (19) & \cite{Jian.2023}\\
        87 & 436373446.4 (14) & 444778798.4	(24)&\\
           &                  & 444778796.11 (4) & \cite{barwood_observation_2003}\\
        86 & 436373122.8 (9)  & 444778474.2	(9)	&\\
           &                  & 444778473.814 (4) & \cite{lybarger_precision_2011}\\ 
        84 & 436372494.1 (14) & 444777846.5	(17)&\\
    \end{tabular}
    \end{ruledtabular}
\end{table}

\subsection{Isotope shifts}
Since multiple measurement schemes were used to extract the same observables, the results were combined into a single value. The final isotope shifts are listed in Tab.\,\ref{tab:shift}. For the D1, D2, and $4D_{\nicefrac{3}{2}}\rightarrow 5P_{\nicefrac{3}{2}}$ transition, the isotope shifts obtained by subtracting the absolute transition frequencies were unified with the results of the direct IS measurements. Agreement is generally within one sigma, with the only exception ($2.5\,\sigma$) being $^{84}$Sr in the $4D_{\nicefrac{3}{2}}\rightarrow 5P_{\nicefrac{3}{2}}$ transition, where we have the least statistics.
For the $4D_{\nicefrac{3}{2}}\rightarrow 5P_{\nicefrac{1}{2}}$ and $4D_{\nicefrac{5}{2}}\rightarrow 5P_{\nicefrac{3}{2}}$ transitions, only acca measurements were performed. Since, apart from the $^{88,86}$Sr pair for the $4D_{\nicefrac{5}{2}}\rightarrow 5P_{\nicefrac{3}{2}}$ transition, we only have acca measurements with the wavemeter, they have the largest uncertainty.

The only isotope shift measurements that are known to us in the D-P manifold are those of $\delta\nu^{86,88}$ and $\delta\nu^{84,88}$ in the $4D_{\nicefrac{3}{2}}\rightarrow 5P_{\nicefrac{1}{2}}$ transition \cite{Dubost.2014}. Our values agree well and reduce the uncertainty by a factor of two. 
Similar to the absolute transition frequencies, we can also calculate the isotope shifts between $^{88}$Sr and the other stable isotopes in the $S_{\nicefrac{1}{2}}\rightarrow D_{\nicefrac{3}{2}, \nicefrac{5}{2}}$ transitions. These are also included in the table and nicely agree with the literature in the two cases in which measurements are available \cite{lybarger_precision_2011, barwood_observation_2003}.

\begin{table*}[t]
\caption{Experimental isotope shifts $\delta\nu_{i}^{A, 88} = \nu_{i}^{A}-\nu_{i}^{88}$ with respect to $^{88}$Sr for all investigated transitions. All values are in MHz. \label{tab:shift}}
    \centering
    \begin{ruledtabular}
    \sisetup{table-format = 4.3(2){$^\dagger$}}
    \begin{tabular}{c*7S@{}}
        A &  {$\delta\nu_{D1}^{A,88}$} & {$\delta\nu_{D2}^{A,88}$ } & {$\delta\nu_{D_{\nicefrac{3}{2}}\rightarrow P_{\nicefrac{1}{2}}}^{A,88}$ } & {$\delta\nu_{D_{\nicefrac{3}{2}}\rightarrow P_{\nicefrac{3}{2}}}^{A,88}$ } & {$\delta\nu_{D_{\nicefrac{5}{2}}\rightarrow P_{\nicefrac{3}{2}}}^{A,88}$ } & {$\delta\nu_{S_{\nicefrac{1}{2}}\rightarrow D_{\nicefrac{3}{2}}}^{A,88}$ } & {$\delta\nu_{S
        _{\nicefrac{1}{2}}\rightarrow D_{\nicefrac{5}{2}}}^{A,88}$ }\\
        \hline
        \noalign{\vskip 0.7mm}
        87 & -56.73 (22)    & -56.45 (23)   & 191.4 (17) & 193.7 (23)   & 189.6 (23) & -248.1 (17)  & -245.7 (25) \\
        86 & -171.53 (21)   & -171.07 (21)  & 400.2 (12) & 400.4 (7)    & 400.2 (19) & -571.8 (12)  & -569.9 (12)\\
        84 & -375.3 (8)     & -375.07 (23)  & 825.2 (18) & 827.1 (21)   & 822.5 (14) & -1200.5 (17) & -1197.6 (18)\\
        \hline
        \noalign{\vskip 0.7mm}
        87 & -59 (7) $^\dagger$     & -56 (3)$^\ast$    &                      & & & & -247.99(4) $^\S$\\
        86 & -170 (3) $^\ddagger$   & -171 (3) $^\ast$  & 402 (2) $^\ddagger$  & & & & -570.264(0) $^\P$\\
        84 & -378 (4) $^\ddagger$   & -373 (5) $^\ast$  & 828 (4) $^\ddagger$  & & & & \\
    \end{tabular}
    \end{ruledtabular}
    {\raggedright
    References: $^\dagger$ \cite{borghs_hyperfine_1983}, $^\ddagger$ \cite{Dubost.2014}, $^\ast$ \cite{Buchinger.1990}, $^\S$ \cite{barwood_observation_2003}, $^\P$ \cite{Manovitz2019}
    \par}
\end{table*}

\begin{figure}[t]
    \centering
    \includegraphics[width=1\linewidth]{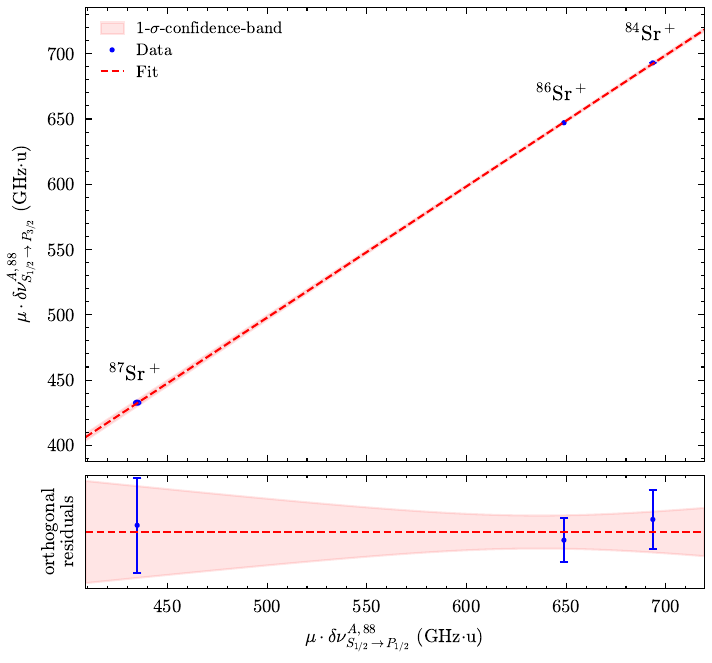}
    \caption{King plot of the modified isotope shifts of the D1 and D2 transitions with $^{88}$Sr as reference. The projected orthogonal residuals are shown below.}
    \label{fig:king_plot_field_shift_ratio}
\end{figure}

Following Eq.\,\eqref{eq:FSR}, plotting the modified isotope shift of the D1 and D2 transition against each other yields the King plot shown in Fig.\,\ref{fig:king_plot_field_shift_ratio}. For the fit, the algorithm described by York \textit{et al}. \cite{York.2004} was used to take uncertainties in both transitions (\textit{i.e.,} in $x$ and $y$ direction) into account in a self-consistent manner. This procedure was performed with each isotope as reference, the plot shown in Fig.\,\ref{fig:king_plot_field_shift_ratio} uses $^{88}$Sr. As can be seen, the fit goes through all data points without any sign of nonlinearity, which is expected at this level of accuracy. Possible nonlinearities arising from off-diagonal hyperfine interactions are expected to be significantly smaller in Sr$^+$ than in neutral Sr \cite{Berengut.2025} due to the absence of closely lying singlet and triplet states. By unifying the slopes of the King plots with different reference isotopes, a final value for the field-shift ratio of $f=1.004(5)$ is obtained. Although the result is still statistically compatible with the non-relativistic limit of 1, it remains fully consistent with the theoretically expected range below the ultra-relativistic upper bound of 1.019.

\subsection{Hyperfine structure constants: experiment and theory}
The HFS constants for $^{87}$Sr were directly fitted to the experimental spectra using a fully unconstrained model in which all hyperfine structure parameters, transition frequencies, a common linewidth, and amplitudes were treated as free parameters. For the $4d\,^2{\!}D_{5/2}\rightarrow5p\,^{2\!}P_{3/2}$ transition, the literature values of Ref.~\cite{barwood_observation_2003} were used as initial values for the fit because of the small hyperfine splittings, while all parameters remained free. As most states are addressed by multiple investigated transitions, for which the agreement lies within two sigma, the results from different transitions are averaged. The final results are listed in Tab.\,\ref{tab:hfs} and agree well with available literature values. While our results cannot compete with the achieved uncertainties in the ground state and states investigated via spectroscopy performed in traps \cite{sunaoshi_precision_1993,barwood_observation_2003}, the excellent agreement provides confidence for those cases in which we report improved or -- to our knowledge -- first data.
\begin{table*}[t]
\caption{Experimental and theoretical HFS parameters for $^{87}$Sr from this work in comparison to literature values from experiment \cite{barwood_observation_2003,sunaoshi_precision_1993,Buchinger.1990} and theory \cite{yu_calculation_2004}. For the theory results from Yu \etal\,, we utilized the same $Q_{I}$ of 0.3102\,b as in our own calculations\label{tab:hfs}.}
    \centering
    \begin{ruledtabular}
        \begin{tabular}{cSSSS}
        {} & {Experiment} & {Experiment} & {Theory} & {Theory} \\
        {} & {This work} & {Ref.} & {This work} & {Ref. \cite{yu_calculation_2004}} \\
        {} & {(MHz)} & {(MHz)} & {(MHz)} & {(MHz)} \\
        \hline     
        \noalign{\vskip 0.7mm}
         $A_{S_{\nicefrac{1}{2}}}$ & -1000.51(4) & -1000.473673 (11) $^\dagger$ & -998(3) & -1003.2 \\
         $A_{P_{\nicefrac{1}{2}}}$ & -177.89 (6) & & -176(1) & -178.4\\
         $A_{P_{\nicefrac{3}{2}}}$ & -36.1 (2) & -36.0(4) $^\ddagger$ & -35.5(5) & -35.1 \\
         $B_{P_{\nicefrac{3}{2}}}$ & 86.4 (6) & 88.5(54) $^\ddagger$ & 85.4(6) & 85.1 (54) \\
         $A_{D_{\nicefrac{3}{2}}}$ & -46.07 (7) & & -46.2(5) & -47.4 \\
         $B_{D_{\nicefrac{3}{2}}}$ & 36.4 (15) & & 35.9(4) & 36.7 (23) \\
         $A_{D_{\nicefrac{5}{2}}}$ & 2.17 (15) & 2.1743 (14) $^\ast$ & 2.25(9) & 2.5 \\
         $B_{D_{\nicefrac{5}{2}}}$ & 49 (4) & 49.11(6) $^\ast$ & 49.1(5) & 52.3(33)  \\
        \end{tabular}
    \end{ruledtabular}
    {\raggedright
    References: $^\dagger$ \cite{sunaoshi_precision_1993}, $^\ddagger$ \cite{Buchinger.1990}, $^\ast$ \cite{barwood_observation_2003}
    \par}
\end{table*}

In order to obtain HFS parameters from theory, we employ the Dirac-Coulomb-Breit interaction Hamiltonian which in atomic units is given by
\begingroup
\setlength{\abovedisplayskip}{4pt}
\setlength{\abovedisplayshortskip}{4pt}
\setlength{\belowdisplayskip}{4pt}
\setlength{\belowdisplayshortskip}{4pt}
\begin{equation}
\begin{split}
H_{\mathrm{at}} \equiv {}&
\sum_i \left[
c\,\vec{\alpha}_i^D\!\cdot\!\vec{p}_i
+(\beta_i^D-1)c^2+V_n(r_i)
\right]
\\[-0.4ex]
&+\sum_{i<j}\left[
\frac{1}{r_{ij}}
-\frac{
\vec{\alpha}_i^D\!\cdot\!\vec{\alpha}_j^D
+(\vec{\alpha}_i^D\!\cdot\!\hat{r}_{ij})
 (\vec{\alpha}_j^D\!\cdot\!\hat{r}_{ij})
}{2r_{ij}}
\right].
\end{split}
\raisetag{2.5ex}
\end{equation}
\endgroup
where $\alpha^D$ and $\beta^D$ are the Dirac matrices, $\vec{p}$ is the momentum operator of a single-particle, $V_n(r)$ represents the nuclear potential seen by an electron and $r_{ij}$ represents the radial difference between the $i^{th}$ and $j^{th}$ electron. Atomic orbitals are generated using Gaussian type orbital functions in the $V^{N-1}$ potential formalism of Dirac-Hartree-Fock (DHF) theory, where $N$ is the total number of electrons of the considered Sr$^+$ ion. To account for the contributions from the residual electron correlation effects, we apply the relativistic coupled-cluster theory at the singles, doubles and triples excitation approximation (RCCSDT method). Details of this method can be found from the earlier studies of  Sr$^+$ properties \cite{Sahoo2006, Chakraborty.2026}.

For $^{87}$Sr, we have used the nuclear magnetic moment $\mu_I= -1.0936030\,\mu_N$ from Ref.\,\cite{Stone2025} and the latest reported nuclear quadrupole moment $Q_I=0.3102\,\text{b}$ from Ref.\,\cite{Chakraborty.2026} to estimate the $A_{\mathrm{hf}}$ and $B_{\mathrm{hf}}$ HFS constants, respectively. 
Corrections from the Bohr-Weisskopf effects are included in the $A_{\mathrm{hf}}$ values in a similar way as discussed in Ref.\,\cite{SahooYb+}. Uncertainties to the calculated values due to incompleteness of basis functions and approximations in the methods are estimated and reported along with the final values in Tab.\,\ref{tab:hfs} .
The calculated HFS constants are within one sigma of the combined uncertainty with our and previous experimental results \cite{Buchinger.1990,barwood_observation_2003,sunaoshi_precision_1993}. 
The sole exception is $A_{P_{\nicefrac{1}{2}}}$, for which it is slightly larger ($\sim 1.5\,\sigma$). We also find close agreement with the relativistic many-body calculations of Yu \etal\ \cite{yu_calculation_2004}. For $A_{P_{\nicefrac{1}{2}}}$, their result is 
closer to the experimental value, but the uncertainty of this value is not stated.

\section{Determination of Leading-order isotope-shift parameters}
\label{sec:LeadOrdISParameters}
We have now established a large set of isotope shifts in all allowed dipole transitions between the $5s$, $5p$, and $4d$ manifolds of fine-structure levels in Sr$^+$ from which we can in principle extract nuclear charge radii. According to Eq.\,\eqref{eq:isotope_shift} this requires knowledge of the atomic parameters $F_i$ and $K_i$, which we obtained in three different ways: 

\begin{enumerate}
    \item[A] (Theory) Mass-shift and field-shift factors from atomic theory. 
    \item[B] (Experimental) King plot according to Eq.\,\eqref{eq:KingPlot2} solely based on charge radii obtained from a combined analysis of muonic-atom and electron-scattering data ($\delta\langle r^2\rangle^{A,A'}_{\mu,e}$) and the measured isotope shift.
    \item[C] (Theory-informed King fit) King fit as in B but with the field-shift factor $F_i$ fixed to the theoretical value as obtained in A to extract an improved mass-shift factor from experiment.
\end{enumerate}
All three procedures are regularly used in literature to extract the atomic parameters, depending on the available information for the respective element.

\subsection{Theory}
\label{sec:Theory}
To calculate the atomic parameters, which are referred to as the first-order isotope shift (IS) parameters ($\langle O^{\mathrm{IS}} \rangle$) for the respective IS operator $O^{\mathrm{IS}}$, we apply the finite-field (FF) approach by expressing the calculated energy of the atomic state in the RCCSDT method with valence orbital $v$ as 
\begin{eqnarray}
E_v(\lambda) = E_v^{(0)} + \lambda \langle O^{IS} \rangle+ \frac{\lambda^2}{2} \langle O^{IS} \rangle^{(2)}  + {\cal O}(\lambda^3)  ,
\label{eqn1}
\end{eqnarray}
where $E_v^{(0)}$ is the energy due to atomic Hamiltonian $H_{at}$, $E_v(\lambda)$ is the total energy due to the modified atomic Hamiltonian $H_{\lambda} = H_{at} + \lambda O^{IS}$ for an arbitrary numerical perturbative parameter $\lambda$. We also apply our recently developed analytical response (AR) approach to estimate these IS parameters by solving the following inhomogeneous equation \cite{SahooAR}
\begin{eqnarray}
(H_{at}-E_v^{(0)}) |\Psi_v^{(1)} \rangle = (\langle O^{IS} \rangle - O^{IS} ) |\Psi_v^{(0)} \rangle ,
\label{eqn2}
\end{eqnarray}
where superscript $0/1$ denotes the order of $\lambda$ in the results. Further descriptions of both the FF and AR approaches in the RCCSDT theory to evaluate the IS constants can be found elsewhere \cite{Sahoo1,Sahoo2,Katyal2025}. The calculated $F$ and $K$ along with their uncertainties of the aforementioned states of Sr$^+$ are presented from both the FF and AR approaches of the RCCSDT method in Tab. \ref{tab:isotope_shift_params}. The agreement with the previously calculated $F_i$ \cite{Munro-Laylim.2022} that are included in the table is very good. 

\subsection{King plot}
\begin{figure*}[t]
    \centering

    \begin{minipage}{0.49\textwidth}
        \centering
        \includegraphics[width=\linewidth]{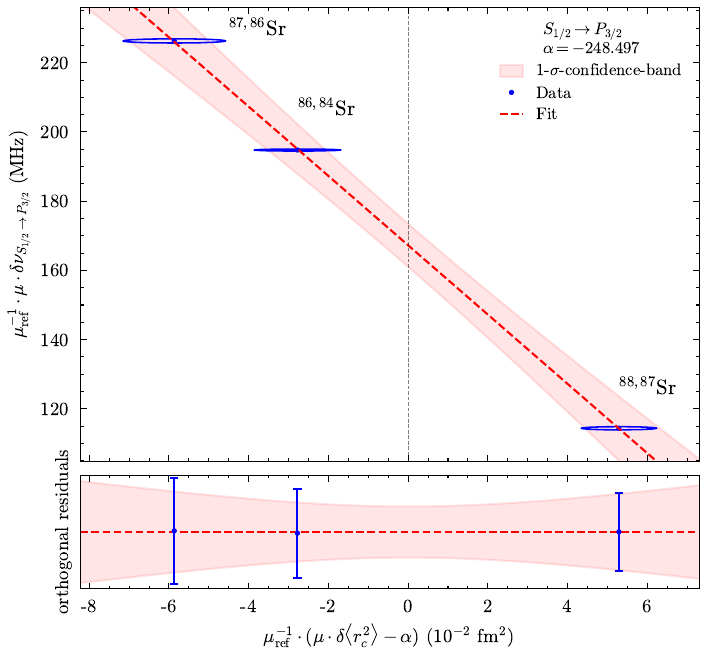}
    \end{minipage}
    \hfill
    \begin{minipage}{0.49\textwidth}
        \centering
        \includegraphics[width=\linewidth]{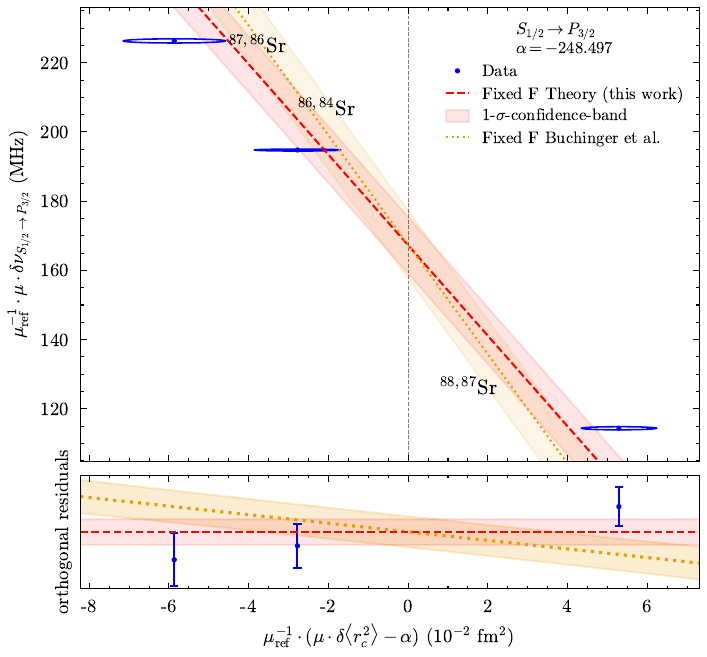}
    \end{minipage}

    \caption{Ladder-type King plots using optical data for the $S_{\nicefrac{1}{2}}\rightarrow P_{\nicefrac{3}{2}}$ transition from this work and the charge radii from literature provided in Tab.\,\ref{tab:input}. \textit{Left:} With free parameters for slope and $y$-intercept utilizing a York-type fitting routine \cite{King.1984, muller_qspec_2025}. \textit{Right:} With the slope fixed to the theoretical value obtained in this work (red) and to that used by Buchinger \etal\ \cite{Buchinger.1990} (yellow). For both plots, both axes were scaled by a reference-pair factor of $\mu_\text{ref}^{-1}=\nicefrac{(M_{88}-M_{86})}{(M_{88}\cdot M_{86})}$ for improved readability.}
    \label{fig:King-plots}
\end{figure*}

The King plot for the D2 transition is shown in the left part of Fig.\,\ref{fig:King-plots} as an example for this approach.
The values of $\delta\langle r_{\mathrm{c}}^2\rangle^{A,A'}_{\mu \mathrm{e}}$ used as input parameters are included in the last line in Tab.\,\ref{tab:input} . They were calculated using the (model-independent) Barrett radii $R_{k\alpha}$ from muonic atom data and the ratio of radial moments $V_2^A$ from elastic electron scattering, both taken from \cite{Fricke.2004} using
\begin{equation}
    \langle r^n_c \rangle^{1/n}_{\mu e}=R^\mu_{k\alpha}/V^{e}_n.
\end{equation}
In the following, we drop the superscripts $\mu$ and $e$ on $R_{k\alpha}$ and $V_2$ and instead indicate the mass number $A$ at this position. $V_2$ is only available for $^{88}$Sr and it was used for all Sr isotopes by Fricke and Heilig \cite{Fricke.2004}. Thus, the authors explicitly report charge-radii differences with purely statistical uncertainties and refrain from providing systematic uncertainties. In order to estimate the effect of the $V_2$-variation in Sr, we estimated $V_2^{A}$ by utilizing experimental data of stable isotopes in close proximity, namely Ge, Zr, Mo and Pd taken from \cite{DeVries.1987, Fricke.2004}. Across each isotopic chain of these elements, the deviation from sphericity $v_2^A=\sqrt{5/3}/V_2^{A}-1$ as defined in \cite{Ohayon.2025}, follows a linear trend towards the nuclear shell closure at $N=50$. Below the shell closure, the trend is falling with increasing mass number, whereas it is rising above. To estimate $V_2^A$ for isotopes other than $^{88}$Sr, we fixed the value of $v_2^A$ for $^{88}$Sr and used the average of the absolute slopes of neighboring elements, yielding $v_2^{A} = v_2^{88} + (88 - A)\cdot \left[-1.2(0.7)\cdot 10^{-4}\right]$, from which $V_2^A$ can then be calculated, see Tab.\,\ref{tab:input}. The uncertainty of the slope was chosen such that all (absolute) slopes obtained in the neighboring elements lie within 1$\sigma$. A plot explaining our procedure can be found in Fig.\,\ref{fig:v2} in Appendix \ref{ap: v2A}.

The differential charge radius can then be calculated by
\begin{align}
\delta\langle r^2_{\mathrm{c}} \rangle^{A,A'} &= \langle r^2_{\mathrm{c}} \rangle^A - \langle r^2_{\mathrm{c}} \rangle^{A'} \nonumber\\
&= \left( \frac{R_{\kappa \alpha}^A}{V_2^A} \right)^2 - \left( \frac{R_{\kappa \alpha}^{A'}}{V_2^{A'}} \right)^2 \nonumber\\
&= \left(\frac{R_{\kappa \alpha}^A}{V_2^A} - \frac{R_{\kappa \alpha}^{A'}}{V_2^{A'}} \right) 
\cdot 
\left(\frac{R_{\kappa \alpha}^A}{V_2^A} + \frac{R_{\kappa \alpha}^{A'}}{V_2^{A'}}\right)\nonumber\\
&=\frac{1}{V_2^{A}}\left(\delta R_{k\alpha}^{A,A'}+\left(1-\frac{V_2^{A}}{V_2^{A'}}\right)R^{A'}_{k\alpha}\right)\nonumber\\
&\cdot\left(\frac{R_{\kappa \alpha}^A}{V_2^A} + \frac{R_{\kappa \alpha}^{A'}}{V_2^{A'}}\right).
\end{align}
In the last step, we utilized the differential Barrett radii between two stable isotopes $\delta R_{k\alpha}^{A,A'}=R_{\kappa \alpha}^A - R_{\kappa \alpha}^{A'}$ provided in \cite{Fricke.2004}, which is known more accurately, as parts of the uncertainties cancel \cite{Fricke.2004, Konig.2021b}. The uncertainty of the $V_2^A$ parameter is often neglected although it has a significant impact on the total uncertainty. Including the uncertainties of the $V_2^A$ as estimated above increases the uncertainties of the differential charge radii linearly up to a maximum factor of 1.8 for the $\delta\langle r_c^2\rangle ^{84, 88}$ pair.

\begin{table*}[t]
    \caption{Input data for the King plots. Masses are taken from the AME \cite{wang_ame_2021} and recent studies of $^{84}$Sr \cite{ge_high-precision_2024}, Barrett radii, changes of the Barrett radii between isotopes, and the $V_2^{88}$ from the book of Fricke and Heilig \cite{Fricke.2004}.
    The $V_2$ ratios of the other isotopes are estimated as described in the text. They are used to extract the rms charge radius $\langle r_{\mathrm{c}}^2 \rangle^{\nicefrac{1}{2}}_A$ according to $ \left\langle r_{\mathrm{c}}^{2}\right\rangle^{\nicefrac{1}{2}}_A = R_{k\alpha}^A/V_2^A$ and the differential ms charge radii $\delta \langle r_{\mathrm{c}}^2 \rangle^{A,88}$ along the chain.}
    \label{tab:input}
    \centering
    \begin{ruledtabular}
    \begin{tabular}{lS[table-format=1.5(3)]S[table-format=1.5(3)]S[table-format=1.5(3)]S[table-format=1.5(3)]}
    {A} & {88} & {87} & {86} & {84} \\
    \hline
    \rule{0mm}{3.5mm}Mass [u] & 87.905612253(6) & 86.908877495(5) & 85.909260725(6) & 83.9134191(13) \\
    Barrett radius $R_{k\alpha}^A$ [fm] & 5.4091(15) & 5.4094(12) & 5.4183(11) & 5.4318(12) \\
    Barrett radius difference $\delta \left({R_{k\alpha}^\mu}\right)^{A,86}$ [10$^{-3}$fm] & -9.2(6) & -8.9(6) & 0 & 13.5(7)\\
    $V_2^A$ & 1.28219 & 1.28203(8) & 1.28187(16) & 1.2816(3)\\
    $\langle r_{\mathrm{c}}^2 \rangle^{\nicefrac{1}{2}}_A$ [fm] & 4.2186(12) & 4.2194(10) & 4.2269(10) & 4.2384(15)\\
    $\delta \langle r^2 \rangle^{A,88}$ [10$^{-3}$fm$^2$] & 0 & 6(6) & 69(6) & 167(11)\\
    \end{tabular}
    \end{ruledtabular}
\end{table*}

To decorrelate the fit results for the slope and $y$-intercept, we can introduce an offset parameter $\alpha$ into Eq.\,\eqref{eq:KingPlot2} 
\begin{equation}
    \mu^{A,A'} \delta \nu^{A,A'} = F_{i} \, \left(\mu^{A,A'}{\delta \langle r^2 \rangle^{A,A'}} - \alpha \right) + K_{i,\, \alpha}
\end{equation}
that shifts the origin of the abscissa into the center of gravity of the data as introduced by Hammen \etal\ \cite{Hammen.2018}.

We also performed a six-dimensional King plot (see Fig. \ref{fig:6d-plot} in Appendix \ref{ap: king-plot}), which simultaneously includes all measured isotope shifts and the reference charge radii; additional details are provided in Appendix \ref{ap: king-plot}. The six-dimensional fit confirms the consistency of our optical data, passing through all data points within one sigma and yielding results consistent with the conventional King plots.

It is important to note that the physically relevant mass-shift factor is obtained by projecting $K_{i,\,\alpha}$ back to $\alpha=0$ to receive $K_{i}\equiv K_{i,\, 0}$, which was done for the results of the King-plots in Tab.~\ref{tab:isotope_shift_params} to compare them to the theory results.  
\begin{table*}[t]
\caption{Atomic factors from the theoretical calculations using the analytical response (AR) and finite field (FF) approach, from experimental data through a King plot procedure, and with a theory-informed King plot as described in Sec.\,\ref{sec:LeadOrdISParameters}. For the latter, we used the finite field results from theory in the King plot. The field-shift factors $F_i$ are in MHz/fm$^2$ and the mass-shift factors $K_i$ are in GHz$\cdot$u.\label{tab:isotope_shift_params}}
\centering
\begin{ruledtabular}
\begin{tabular}{c S S S S S}
{} & \multicolumn{2}{c}{Theory (this work)} & {Theory} & {Experimental} & {Theory-informed} \\
{} & AR & FF & {Ref. \cite{Munro-Laylim.2022}} & {} & {} \\
\cline{2-3}
\hline
F$_{S_{\nicefrac{1}{2}}\rightarrow P_{\nicefrac{1}{2}}}$ & -1300(10) & -1301(10) & -1331 & -997(130) & -1301(10) \\
F$_{S_{\nicefrac{1}{2}}\rightarrow P_{\nicefrac{3}{2}}}$ & -1307(5) & -1308(5) & -1340 & -1001(131) & -1308(5) \\
F$_{D_{\nicefrac{3}{2}}\rightarrow P_{\nicefrac{1}{2}}}$ & 275(15) & 265(10) & 282 & 221(52) & 265(10) \\
F$_{D_{\nicefrac{3}{2}}\rightarrow P_{\nicefrac{3}{2}}}$ & 267(15) & 258(10) & 273 & 161(59) & 258(10) \\
F$_{D_{\nicefrac{5}{2}}\rightarrow P_{\nicefrac{3}{2}}}$ & 259(15) & 246(12) & 263 & 264(69) & 246(12) \\
F$_{S_{\nicefrac{1}{2}}\rightarrow D_{\nicefrac{3}{2}}}$ & -1574(25) & -1566(20) & -1613 & -1219(165) & -1566(20) \\
F$_{S_{\nicefrac{1}{2}}\rightarrow D_{\nicefrac{5}{2}}}$ & -1566(20) & -1554(15) & -1602 & -1270(177) & -1554(15) \\
\hline
K$_{S_{\nicefrac{1}{2}}\rightarrow P_{\nicefrac{1}{2}}}$ & 237(50) & 235(50) &  & 386(40) & 310(31) \\
K$_{S_{\nicefrac{1}{2}}\rightarrow P_{\nicefrac{3}{2}}}$ & 231(50) & 231(50) &  & 384(40) & 308(31) \\
K$_{D_{\nicefrac{3}{2}}\rightarrow P_{\nicefrac{1}{2}}}$ & -1340(80) & -1394(50) &  & -1457(17) & -1444(9) \\
K$_{D_{\nicefrac{3}{2}}\rightarrow P_{\nicefrac{3}{2}}}$ & -1346(80) & -1397(50) &  & -1480(21) & -1450(10) \\
K$_{D_{\nicefrac{5}{2}}\rightarrow P_{\nicefrac{3}{2}}}$ & -1359(80) & -1393(50) &  & -1437(22) & -1442(11) \\
K$_{S_{\nicefrac{1}{2}}\rightarrow D_{\nicefrac{3}{2}}}$ & 1580(100) & 1628(80) &  & 1842(51) & 1756(38) \\
K$_{S_{\nicefrac{1}{2}}\rightarrow D_{\nicefrac{5}{2}}}$ & 1590(100) & 1624(80) &  & 1821(55) & 1749(38) \\
\end{tabular}
\end{ruledtabular}
\end{table*}

Our experimental King fit for the $S_{\nicefrac{1}{2}}\rightarrow P_{\nicefrac{3}{2}}$ transition yields $F=-1001(131)$\,MHz/fm$^2$ and $K=384(40)$\,GHz$\cdot$u, in good agreement with the values $F=-1020(140)$\,MHz/fm$^2$ and $K=412(42)$\,GHz$\cdot$u reported by Fricke and Heilig \cite{Fricke.2004}. Their smaller uncertainties were obtained assuming the same $V_2$ value for all isotopes without uncertainties. Using the methods and input values documented in their compilation, we were not able to reproduce the quoted uncertainties.

There are surprisingly large differences of up to 2\,$\sigma$ in the field-shift factors between theory and experiment, even though the theoretical accuracy is assumed to be at the 1\% level and the linearity of the King plot is, in most cases, excellent. This is particularly critical with respect to the D1 and D2 transitions, which are most sensitive to the charge radius and exhibit a discrepancy of about 30\%. This is drastically different from the case of Mg$^+$, where an excellent agreement between theory and King-fit result was obtained \cite{Yordanov.2012, sahoo_accurate_2010}. For Sr$^+$, good agreement is only found for the $D_{\nicefrac{5}{2}}\rightarrow P_{\nicefrac{3}{2}}$ transition, which has a comparably low sensitivity.

\subsection{Theory-informed King plot}
We have performed a theory-informed King plot in which the field-shift factor $F$ is fixed to its theoretical value in order to obtain an improved mass-shift factor $K$, which is generally considered less reliable from theory. The theoretical uncertainty of $F$ is propagated using a Monte Carlo procedure: In each realization, $F$ is sampled from a Gaussian distribution defined by its theoretical central value and uncertainty, and $K$ is refitted while accounting for the experimental uncertainties in both King-plot coordinates. The uncertainty reported for $K$ is obtained from the resulting distribution and therefore contains contributions from both the experimental data and the theoretical uncertainty of $F$. The correlation between $F$ and $K$ is retained when propagating the result to derived quantities. The quoted $\chi^2$ value, however, is evaluated with $F$ fixed to its central theoretical value.

This theory-informed approach is sometimes used when only two or three stable isotopes exist and a reliable field-shift calculation is available, as is the case for alkaline atoms (odd $Z$) with their comparatively ``simple'' atomic structure \cite{konig_nuclear_2024}. The result of the theory-informed King fit for the $S_{\nicefrac{1}{2}}\rightarrow P_{\nicefrac{3}{2}}$ transition is shown in the right panel of Fig.\,\ref{fig:King-plots} and compared with the free King fit in the left panel. Fixing the slope to the central theoretical value increases $\chi^2$ from 0.006 to 3.21, as is clearly visible in the figure. We additionally performed a fit with $F$ restricted to the value previously used by Buchinger \etal\ \cite{Buchinger.1990}, which was obtained using the semi-empirical Goudsmit--Fermi--Segrè approach. This results in an even larger value of $\chi^2=7.92$, represented by the dotted line in the right panel of Fig.\,\ref{fig:King-plots}.

These surprisingly large deviations may arise from several sources, and it is important to distinguish between the calculations of the field-shift ($F$) and mass-shift ($K$) factors. Calculations of the specific mass shift are known to be notoriously difficult, as they are highly sensitive to electron-correlation effects and challenging to converge reliably. In particular, missing higher-order correlation contributions, such as quadruple excitations, may significantly affect the calculated $K$ values and their uncertainties. In this context, the observed deviations of the mass-shift factors, which remain within $2\sigma$ of the combined uncertainties, are not alarming. By contrast, calculations of the field-shift factors are generally considered more robust, with different theoretical approaches often agreeing at the level of a few percent \cite{Cheal2012,Filippin2016,Ohayon2022}. Consequently, the discrepancies observed for the $F$ factors are considerably more surprising.

Missing QED contributions may account for part of these discrepancies, particularly for transitions involving the relativistic $S_{\nicefrac{1}{2}}$ and $P_{\nicefrac{1}{2}}$ states, but are unlikely to explain their full magnitude. For the mass-shift factors, the noticeable differences between results obtained using the non-relativistic and relativistic mass-shift operators indicate that relativistic corrections are relevant, while higher-order electron-correlation effects may provide additional contributions.

A second possible source of systematic effects originates from the nuclear input data used in the King-fit procedure. While the optical isotope-shift measurements presented here have very small experimental uncertainties, the extraction of atomic factors relies on differential charge radii derived by combining Barrett radii from muonic-atom spectroscopy with ratios of radial moments $V_2^A$ obtained from elastic electron scattering.

More generally, recent reevaluations of muonic-atom data have demonstrated that improved treatments of QED and nuclear-structure effects can lead to noticeable changes in the extracted nuclear charge radii \cite{sun_208pb_2025}. Such efforts are now being extended to other nuclei, including $^{90}$Zr \cite{Beyer2025}, and are supported by recently developed theoretical frameworks for medium-mass muonic atoms \cite{Rathi2026}. For the stable Sr isotopes compiled by Fricke and Heilig \cite{Fricke.2004}, however, the quoted uncertainties of both the Barrett equivalent radii and the differential Barrett moments are dominated by the theoretical uncertainty of the nuclear-polarization correction rather than by the experimental uncertainties of the measured muonic transition energies. Consequently, a modern reevaluation of the theoretical corrections applied to the published experimental energies appears considerably more promising than a reassessment of the experimental energies themselves.

The dominant theoretical uncertainty in the extraction of the Barrett radii arises from the nuclear-polarization (NPOL) correction \cite{Gorchtein.2026}. In the analysis summarized by Fricke and Heilig \cite{Fricke.2004}, uncertainties corresponding to 30\% of the correction were assigned to the absolute Barrett radii, while differential Barrett moments were assigned uncertainties of 10\% of the larger correction of the isotope pair. In the absence of more rigorous calculations, we follow the same prescription in the present work. Modern calculations of the nuclear-polarization contribution, including improved treatments of nuclear correlations and odd-mass nuclei, could improve both the reliability of the extracted charge radii and the associated uncertainty estimates.

The conversion of the Barrett radii into rms charge radii additionally depends on the isotope dependence of the ratio of radial moments $V_2^A$ obtained from elastic electron scattering. Since only $V_2^{88}$ is available for Sr, the isotope dependence had to be estimated from neighboring isotopic chains. As shown in Fig.\,\ref{fig:v2}, this estimate was derived primarily from even--even nuclei. However, for the Sn isotopic chain, where $V_2^A$ values are also available for odd-$A$ isotopes, these values follow the approximately linear trend between their even--even neighbors \cite{Fricke.2004}. This observation suggests that odd--even effects in $V_2^A$ are small, although anomalous behavior of $^{87}$Sr cannot be excluded entirely. In any case, the estimated variation in $V_2^A$ appears too small to fully explain the observed discrepancies.

\section{Differential charge radii}
\label{sec:ChargeRadii}
In this section, we first investigate the consistency of the differential ms charge radii of the stable isotopes based on the different transitions and then the effect of the observed differences in atomic factors onto the charge radii extracted for the whole chain of Sr isotopes.

\subsection{Stable isotopes in all transitions}

Using the obtained uncorrelated atomic factors and precise isotope shift measurements, we calculate the differential charge radii according to
\begin{equation}
    \delta \langle r{_\mathrm{c}}^2 \rangle^{A,A'} = \frac{\delta \nu^{A,A'} - \frac{K_{i,\,\alpha}}{\mu^{A,A'}}}{F_i} + \frac{\alpha}{\mu^{A,A'}}. 
    \label{eq:charge_radius_extraction}
\end{equation}
Thus for the experimental data, we do not use the $\alpha=0$ projected values of atomic factors from Tab.\,\ref{tab:isotope_shift_params}, but the full sets with $\alpha\neq0$ and $K_{\alpha, i}$ given in Tab.\,\ref{Tab:different_King_plots} in Appendix\,\ref{ap: king-plot}. For the theory values, we utilized the values given in Tab. \ref{tab:isotope_shift_params} with $\alpha=0$. The results for the individual transitions and the three pairs of atomic factors are provided in Tab.\,\ref{tab:radii}. 
The average value of $\delta \langle r_{\mathrm{c}}^2\rangle^{A, 88}$ provided for each set of isotope-shift parameters and each isotope $A=87,\,86,\,84$ is the weighted average of all transitions and its uncertainty has been chosen as the smallest obtained from an individual transition, with no further reduction, since they are all correlated through the King plot with the same $\mu^{A,A'}\,\delta\langle r_{\mathrm{c}}^2\rangle^{A, 88}$ data. The same approach was used for the theory data. As expected given the disagreement in the atomic factors, the charge radii calculated using the theoretical and experimental methods differ by up to $2\sigma$. However, the theory-informed King plot yields good agreement with the experimental results.

\begin{figure}[t]
    \centering
    \includegraphics[width=1\linewidth, clip=true, trim=0cm 0cm 0cm 0mm]{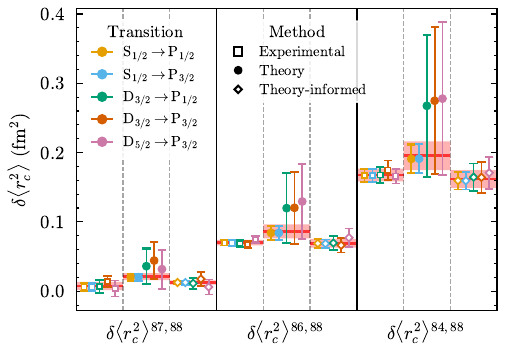}
    \caption{Differential ms nuclear charge radii of $^{87}$Sr (left), $^{86}$Sr (middle) and $^{84}$Sr (right) with respect to $^{88}$Sr as obtained from the isotope-shift data measured in this work (see Tab.\,\ref{tab:radii}). The five investigated transitions are color coded and the three parameter sets ($F_i$, $K_i$) from the respective free King-plot ($\square$), theory ($\bullet$), and the theory-informed King plot with fixed $F_i$ ($\diamond)$ use different markers. The weighted average of the five transitions for each isotope and parameter set is represented by the corresponding horizontal line. The shaded region indicates the estimated uncertainty.}
    \label{fig:new_radii}
\end{figure}

The resulting charge radii obtained from the three different methods and the individual transitions listed in Tab.\,\ref{tab:radii} are shown in Fig.\,\ref{fig:new_radii}. In general, the approaches based on King-plot analyses yield significantly more consistent charge radii than the direct use of the theoretical ($K_i$, $F_i$) parameter sets, for which a substantially larger scatter is observed. The D1 and D2 transitions, where the calculations yielded the highest accuracy (see Tab.\,\ref{tab:isotope_shift_params}), exhibit the smallest uncertainties, whereas the uncertainties for the $4d\rightarrow 5p$ transitions are considerably larger.

Although the quoted uncertainties broadly reflect the observed scatter, the charge radii extracted from the purely theoretical approach are systematically larger than those obtained from the experimentally constrained King-plot analyses. This may indicate correlated systematic effects in the calculations. Since all transitions were calculated within a similar computational framework, missing contributions such as larger basis sets, QED corrections to the mass-shift, or higher-order electron-correlation effects including quadruple excitations could shift the calculated parameters coherently across all transitions, leading to a systematic offset in the extracted charge radii.

Constraining the King-plot analysis with theoretical $F_i$ factors largely removes this discrepancy. Even for the D1 and D2 transitions, where the theoretical and experimental field-shift factors differ by approximately 30\%, the averaged charge radii from all transitions remain consistent within the quoted uncertainties.

\begin{table*}[t]
    \caption{Differential ms nuclear charge radii $\delta \langle r_{\mathrm{c}}^2 \rangle^{A,88}$ of the stable Sr isotopes obtained from the experimental isotope shifts of this work in all five investigated transitions using the three sets of isotope-shift parameters ($K_i$, $F_i$) derived from theory, a free King plot and a theory-informed King plot as listed in Tab.\,\ref{tab:isotope_shift_params} and described in the text. At the top are reference values calculated directly from the Barrett radii given by Fricke and Heilig \cite{Fricke.2004} and our estimated values for $V_2^A$ (see Tab.\,\ref{tab:input}) and a literature value from Buchinger \etal \cite{Buchinger.1990}, where isotope shift measurements were combined with theory values from a Goudsmit-Fermi-Segrè (GFS) approach. All values are in $10^{-3}$\,fm$^2$.  \label{tab:radii}}
\centering
\begin{ruledtabular}
\begin{tabular}{lSSS}
{ } & {$\delta \langle r_{\mathrm{c}}^2 \rangle^{87,88}$} & {$\delta \langle r_{\mathrm{c}}^2 \rangle^{86,88}$} & {$\delta \langle r_{\mathrm{c}}^2 \rangle^{84,88}$ } \\
\hline
From Barrett radii \cite{Fricke.2004} & 6(6) & 69(6) & 167(11) \\
From IS and GFS \cite{Buchinger.1990} & 7(4) & 50(8) & 116(16) \\
\hline
\vspace{-1mm} \\
\multicolumn{4}{l}{\text{$F_i$ \& $K_i$ from King plot\rule{0mm}{3.5mm}}} \\
\hline
\noalign{\vskip 0.7mm}
$S_{\nicefrac{1}{2}}\rightarrow P_{\nicefrac{1}{2}}$ & 6(5) & 70(6) & 167(13) \\
$S_{\nicefrac{1}{2}}\rightarrow P_{\nicefrac{3}{2}}$ & 6(5) & 69(6) & 167(13) \\
$D_{\nicefrac{3}{2}}\rightarrow P_{\nicefrac{1}{2}}$ & 6(12) & 69(12) & 167(22) \\
$D_{\nicefrac{3}{2}}\rightarrow P_{\nicefrac{3}{2}}$ & 4(21) & 57(17) & 162(31) \\
$D_{\nicefrac{5}{2}}\rightarrow P_{\nicefrac{3}{2}}$ & 8(13) & 77(13) & 170(22) \\
\hline
Average (King plot)\rule{0mm}{3.5mm} & 6(5) & 70(6) & 167(13) \\
\hline
\vspace{-1mm} \\
\multicolumn{4}{l}{\text{$F_i$ \& $K_i$ from Theory\rule{0mm}{3.5mm}}} \\
\hline
\noalign{\vskip 0.7mm}
$S_{\nicefrac{1}{2}}\rightarrow P_{\nicefrac{1}{2}}$ & 20(5) & 84(10) & 191(21) \\
$S_{\nicefrac{1}{2}}\rightarrow P_{\nicefrac{3}{2}}$ & 20(5) & 84(10) & 191(21) \\
$D_{\nicefrac{3}{2}}\rightarrow P_{\nicefrac{1}{2}}$ & 36(25) & 120(50) & 267(103) \\
$D_{\nicefrac{3}{2}}\rightarrow P_{\nicefrac{3}{2}}$ & 44(27) & 120(52) & 275(106) \\
$D_{\nicefrac{5}{2}}\rightarrow P_{\nicefrac{3}{2}}$ & 32(28) & 129(55) & 278(111) \\
\hline
Average (Theory)\rule{0mm}{3.5mm} & 21(5) & 86(10) & 195(21) \\
\hline
\vspace{-1mm} \\
\multicolumn{4}{l}{\text{$F_i$ from Theory, $K_i$ from theory-informed King plot\rule{0mm}{3.5mm}}} \\
\hline
\noalign{\vskip 0.7mm}
$S_{\nicefrac{1}{2}}\rightarrow P_{\nicefrac{1}{2}}$ & 12(3) & 69(6) & 159(13) \\
$S_{\nicefrac{1}{2}}\rightarrow P_{\nicefrac{3}{2}}$ & 12(3) & 68(6) & 159(13) \\
$D_{\nicefrac{3}{2}}\rightarrow P_{\nicefrac{1}{2}}$ & 11(8) & 69(10) & 164(20) \\
$D_{\nicefrac{3}{2}}\rightarrow P_{\nicefrac{3}{2}}$ & 18(10) & 66(11) & 164(22) \\
$D_{\nicefrac{5}{2}}\rightarrow P_{\nicefrac{3}{2}}$ & 6(11) & 77(13) & 170(23) \\
\hline
Average (Theory-informed) \rule{0mm}{3.5mm} & 12(3) & 69(6) & 162(13) \\
\end{tabular}
\end{ruledtabular}
\end{table*}

\subsection{Differential ms charge radii of $^{77-100}$Sr}

The isotope shifts of the D2 transition previously measured for the radioactive isotopes $^{78-100}$Sr \cite{Buchinger.1990} provide a sensitive illustration of the consequences of the different atomic factors discussed above. We recalculate the corresponding differential charge radii using the field- and mass-shift factors obtained from the experimental King plot, from atomic theory, and from the theory-informed King fit. The purpose of this comparison is not to establish three independent sets of recommended charge radii, but to quantify how the unresolved discrepancy between the atomic factors propagates along the radioactive isotope chain.

Buchinger \etal\,\cite{Buchinger.1990} used a field-shift factor of $F_{\mathrm{D2}}=-1582(49)\,$MHz/fm$^2$, obtained with the semi-empirical Goudsmit--Fermi--Segrè approach. This value is approximately 50\% larger in magnitude than our experimental King-plot result and 20\% larger than our theoretical value, as also illustrated in Fig.\,\ref{fig:King-plots}. Buchinger \etal\ additionally reported a relativistic Dirac--Fock value of $F_{\mathrm{D2}}=-1174$\,MHz/fm$^2$, but considered the semi-empirical result more reliable and used it to extract the differential charge radii. The small uncertainty assigned to their field-shift factor corresponds to an estimated accuracy of 2\% for the semi-empirical approach, whereas uncertainties of approximately 20\% are now considered more realistic for such estimates \cite{Ohayon.2025}.

\begin{figure}[!htbp]
    \centering
    \includegraphics[width=\linewidth]{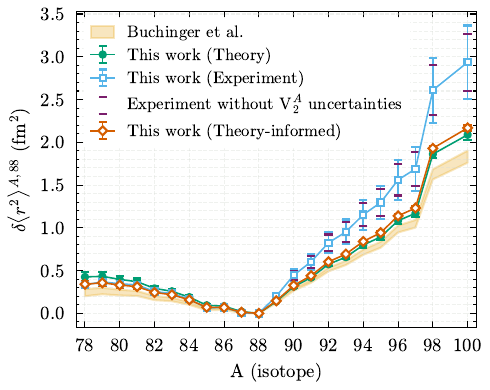}
    \caption{The differential mean-square nuclear charge radii along the Sr isotope chain, relative to $^{88}$Sr, to the choice of atomic factors. All results are based on the D2 isotope shifts measured by Buchinger \etal\ \cite{Buchinger.1990}. Shown are the original evaluation by Buchinger \etal, the results obtained with the theoretical atomic factors, the free experimental King fit, and the theory-informed King fit. The additional caps indicate the uncertainty of the experimental King-plot result when the $V_2^A$ values are treated as exact.}
    \label{fig:sr_chain}
\end{figure}

The resulting differential charge radii are compared in Fig.\,\ref{fig:sr_chain}. Constraining the field-shift factor $F$ in the theory-informed King fit reduces the slope--intercept correlation and, consequently, the uncertainty of the fitted mass-shift factor $K$. This leads to smaller uncertainties than in the free King-plot analysis, while the resulting radii remain close to those obtained with the fully theoretical atomic factors.

The radii obtained with the different approaches are incompatible within their quoted uncertainties, particularly for the neutron-rich isotopes beyond the $N=50$ shell closure. The discrepancies are smaller on the neutron-deficient side but increase progressively towards $^{78}$Sr. The uncertainties shown in Fig.\,\ref{fig:sr_chain} are conditional on the input data and assumptions of each individual approach; they do not include the spread between the approaches. Accordingly, we do not adopt any of these results as a recommended set of differential charge radii. Rather, Fig.\,\ref{fig:sr_chain} demonstrates an unresolved method dependence in the extraction of the Sr charge radii.

At present, the origin of this inconsistency cannot be identified unambiguously. Our RCCSDT field-shift factors agree well with previous calculations based on relativistic Hartree--Fock methods combined with many-body perturbation theory and the random-phase approximation \cite{Munro-Laylim.2022}. The differences among these calculations are small compared with the approximately 30\% discrepancy between theory and our experimental King-plot result (Tab.\,\ref{tab:isotope_shift_params}). At the same time, constraining the King-plot slope to the theoretical field-shift factor substantially increases the $\chi^2$ of the fit. The stable Sr isotopes span only small changes in their mean-square charge radii, making the extracted King-plot slope particularly sensitive to small changes in the nuclear input. These inputs rely on the theoretical interpretation of historical muonic-atom data, including nuclear-polarization corrections, together with the conversion of Barrett radii using the isotope-dependent $V_2^A$ factors. The present comparison therefore does not establish whether the discrepancy originates predominantly from the atomic calculations or from the nuclear input to the experimental King plot.

Resolving this discrepancy requires improved information on both the atomic and nuclear sides. As discussed above, a modern reevaluation of the theoretical interpretation of the published muonic-atom data, particularly through improved QED and nuclear-polarization corrections, could provide revised Barrett radii. Measuring several transitions could better constrain the nuclear charge distributions, while including the long-lived isotope $^{90}$Sr would be particularly valuable because it lies beyond the $N=50$ shell closure, where the charge radii change more strongly.

The discrepancy is also relevant beyond Sr. Sr$^+$, with one valence electron outside a closed-shell core, is generally considered a favorable system for high-accuracy atomic-structure calculations. A difference of the present magnitude therefore motivates similarly stringent comparisons in other monovalent systems for which independent nuclear-radius information and a sufficient number of stable isotopes are available. Such benchmarks are particularly important because charge-radius extractions for elements without sufficient stable isotopes must rely directly on calculated atomic factors. The present analysis also illustrates the importance of publishing the isotope shifts, atomic factors, nuclear input data, and relevant correlations in sufficient detail to permit subsequent re-evaluations when improved atomic or nuclear information becomes available \cite{Angeli.2026}.

\begin{table*}
    \caption{Reevaluated differential ms nuclear charge radii $\delta\langle r_{\mathrm{c}}^2\rangle^{A,88}$ for radioactive Sr isotopes based on the isotope shifts $\delta\nu^{A, 88}$ in the $5s\,^2\!S_{\nicefrac{1}{2}}\rightarrow 5p\,^2\!P_{\nicefrac{3}{2}}$ transition measured by Buchinger \etal\ \cite{Buchinger.1990}. The three different methods disagree for heavier masses, as can also be seen in Fig.\,\ref{fig:sr_chain}.}
    \label{tab:radio_radii}
    \centering    
    \begin{ruledtabular}
    \renewcommand{\arraystretch}{1.15}
    \begin{tabular}{
      l
       S[table-format=4.0(2)]
       S[table-format=1.3(2)]
       S[table-format=1.3(2)]
       S[table-format=1.3(2)]
       S[table-format=1.3(2)]
    }
    A & {$\delta\nu^{A, 88}$} & \multicolumn{4}{c}{\shortstack{$\delta\langle r_{\mathrm{c}}^2\rangle^{A, 88}$}} \\
    \cline{3-6}
    \noalign{\vskip 0.7mm}
     & {[Ref]} & {[Ref]} & \multicolumn{3}{c}{This work} \\
    \cline{4-6}
    \noalign{\vskip 0.7mm}
    {} & {\cite{Buchinger.1990}} & {\cite{Buchinger.1990}} & {Exp.} & {Theo.} & {Theo.-inf.}\\
    \hline
    \noalign{\vskip 0.7mm}
    78 & -893(12) & 0.24(4) & 0.33(4) & 0.43(6) & 0.34(4) \\
    79 & -865(10) & 0.26(4) & 0.37(3) & 0.43(5) & 0.36(3) \\
    80 & -782(11) & 0.24(3) & 0.35(3) & 0.40(4) & 0.33(3) \\
    81 & -710(9) & 0.23(3) & 0.33(3) & 0.37(4) & 0.31(2) \\
    82 & -574(10) & 0.18(2) & 0.25(2) & 0.29(3) & 0.24(2) \\
    83 & -495(6) & 0.16(2) & 0.232(19) & 0.26(3) & 0.217(17) \\
    {83m} & -456(7) & 0.13(2) & 0.193(18) & 0.23(3) & 0.187(17) \\
    84 & -373(5) & 0.116(16) & 0.165(14) & 0.19(2) & 0.158(13) \\
    85 & -216(5) & 0.048(12) & 0.062(12) & 0.094(16) & 0.071(10) \\
    {85m} & -259(5) & 0.075(12) & 0.105(11) & 0.127(16) & 0.104(10) \\
    86 & -171(3) & 0.050(8) & 0.069(7) & 0.084(10) & 0.068(7) \\
    87 & -56(3) & 0.007(4) & 0.006(6) & 0.020(5) & 0.012(4) \\
    {87m} & -7(4) & 0.024(5) & -0.043(11) & -0.018(6) & -0.025(4) \\
    88 & 0 & 0 & 0 & 0 & 0 \\
    89 & -152(2) & 0.124(5) & 0.20(3) & 0.139(5) & 0.146(3) \\
    90 & -349(6) & 0.277(13) & 0.45(7) & 0.312(11) & 0.326(8) \\
    91 & -460(5) & 0.374(16) & 0.60(9) & 0.418(15) & 0.440(10) \\
    92 & -636(8) & 0.51(2) & 0.83(12) & 0.57(2) & 0.603(14) \\
    93 & -718(7) & 0.59(3) & 0.95(15) & 0.66(2) & 0.693(16) \\
    94 & -876(9) & 0.71(3) & 1.15(18) & 0.80(3) & 0.841(19) \\
    95 & -975(8) & 0.80(3) & 1.3(2) & 0.89(3) & 0.94(2) \\
    96 & -1198(10) & 0.99(4) & 1.6(2) & 1.08(4) & 1.14(2) \\
    97 & -1285(11) & 1.05(4) & 1.7(3) & 1.17(4) & 1.23(3) \\
    98 & -2166(11) & 1.62(6) & 2.6(4) & 1.86(5) & 1.93(3) \\
    100 & -2420(20) & 1.83(7) & 2.9(4) & 2.09(6) & 2.17(4) \\
  \end{tabular}
\end{ruledtabular}
\end{table*}

\section{\label{sec:Summary}Summary}

We performed collinear laser spectroscopy of the $5s\rightarrow 5p$ and the $4d\rightarrow5p$ transitions on all stable Sr$^+$ isotopes and achieved accuracies down to 600\,kHz and 210\,kHz for transition frequencies and isotope shifts, respectively. We also extracted HFS parameters for $^{87}$Sr, which were previously experimentally unknown for the $P_{\nicefrac{1}{2}}$ and $D_{\nicefrac{3}{2}}$ states. For the other states, they are in excellent agreement with literature values and the accuracy of $B(P_{\nicefrac{3}{2}})$ has been improved by an order of magnitude. All values agree reasonably well with our theoretical results.

A central focus of this work was the determination of the field- and mass-shift factors $F$ and $K$. Our coupled-cluster calculations differ by up to $2.3\,\sigma$ from the values extracted through a King-plot analysis using our optical data and nuclear charge radii derived from muonic-atom spectroscopy and elastic electron scattering \cite{Fricke.2004}. For the D1 and D2 transitions, which have the largest sensitivity to changes in the nuclear charge radius, the field-shift factors differ by approximately 30\%. A theory-informed King fit, in which $F$ was constrained to the theoretical result while $K$ was determined from the stable-isotope data, reduced the uncertainty of $K$ but resulted in a substantially larger $\chi^2$ than the free King fit. The present analysis does not allow us to determine whether the discrepancy originates predominantly from the atomic calculations or from the nuclear input used in the experimental King plot.

Reliable atomic factors are required for applications ranging from the extraction of the Weinberg angle from parity-nonconservation measurements in Cs \cite{Dzuba.2012,Sahoo2021,Roberts2022,Sahoo2022,Flambaum2026} to the determination of nuclear charge radii from isotope shifts of short-lived isotopes \cite{Kreim2014,Koszorus.2021,Katyal2025} and proposed charge-radius measurements using x-ray transitions in Na-like highly charged ions \cite{Silwal2025,Staiger2025,Hosier2025}. The discrepancy found in Sr$^+$, a nominally favorable system for high-accuracy atomic-structure calculations, therefore motivates similarly stringent comparisons in other monovalent systems. In particular, precise isotope shifts are already available for Ba$^+$ \cite{Imgram.2019}, making corresponding atomic-structure calculations in this system highly desirable.

Finally, we evaluated how the different atomic factors affect the differential charge radii extracted from the D2 isotope shifts of $^{78-100}$Sr measured by Buchinger \etal\,\cite{Buchinger.1990}. The resulting radii show a substantial method dependence and become incompatible within their quoted uncertainties, particularly beyond the $N=50$ shell closure. These uncertainties are conditional on the input data and assumptions of the respective approaches and do not account for the spread between them. We therefore do not recommend any of the resulting datasets as a definitive set of Sr charge radii. Improved atomic-structure calculations together with a modern reevaluation of the QED and nuclear-polarization corrections used in the analysis of the existing muonic-atom data, and new measurements including $^{90}$Sr would help to resolve this inconsistency.

\begin{acknowledgments}
We acknowledge support by the Deutsche Forschungsgemeinschaft (DFG, German Research Foundation) -- Project-ID 279384907 -- SFB 1245 and by the BMFTR under Contract Nos. 05P21RDFN1 and 05P24RD8, and Fonds Wetenschappelijk Onderzoek (FWO, Belgium) Odysseus Project No. G0F7321N. J.P., H.B. and I.L. acknowledge support from HGS-HIRe. The work of B.K.S. at the Physical Research Laboratory (PRL) is supported by ANRF with grant no. CRG/2023/002558 and the Department of Space, Government of India. Atomic calculations were performed on the ParamVikram-1000 HPC cluster at PRL, Ahmedabad, Gujarat, India.

\end{acknowledgments}

\appendix
\section{Estimation of the ratio of radial moments $V_2^A$} \label{ap: v2A}
\begin{figure*}[t]
    \centering
    \includegraphics[width=\linewidth]{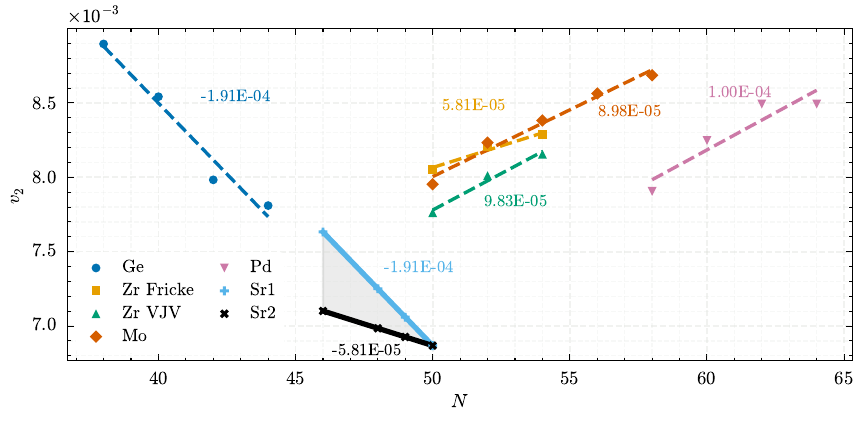}
    \caption{The deviation from sphericity plotted against the neutron number $N$ for different isotopes near $N=50$. The two Sr lines are our lower and upper estimate.}
    \label{fig:v2}
\end{figure*}
In Fricke and Heilig \cite{Fricke.2004} only $V_2^{88}$ is given. While they use this value without uncertainty for all isotopes in their King plot, we know that this is physically incorrect. We chose to use the linear trends of nearby elements towards the neutron shell close $N=50$ as an estimate for this value. In Fig.\,\ref{fig:v2}, the deviation from sphericity $v_2^A=\sqrt{5/3}/V_2^{A}-1$ is plotted for the elements of Ge(Z=32) \cite{Fricke.1995, Mallot.1985, DeVries.1987}, Zr(Z=40) \cite{Fricke.1995, dreher_isotopic_1975, DeVries.1987}, Mo(Z=42) \cite{Fricke.1995, dreher_isotopic_1975, DeVries.1987} and Pd(Z=46) \cite{Fricke.1995, laan_electron_1986, DeVries.1987}. There are two conflicting sets of values plotted for Zr. Across all of these sets, we performed linear fits. We then took the largest and smallest absolute values for the slopes as an upper and lower boundary for the linear trend from the fixed value of $V_2^{88}$. We then took the average between those boundaries as the value used in our calculations.

\section{King Plots in detail: Experimental, Theory-informed and 6-D King Plot} \label{ap: king-plot}
\begin{figure*}
    \centering
    \includegraphics[width=1\linewidth]{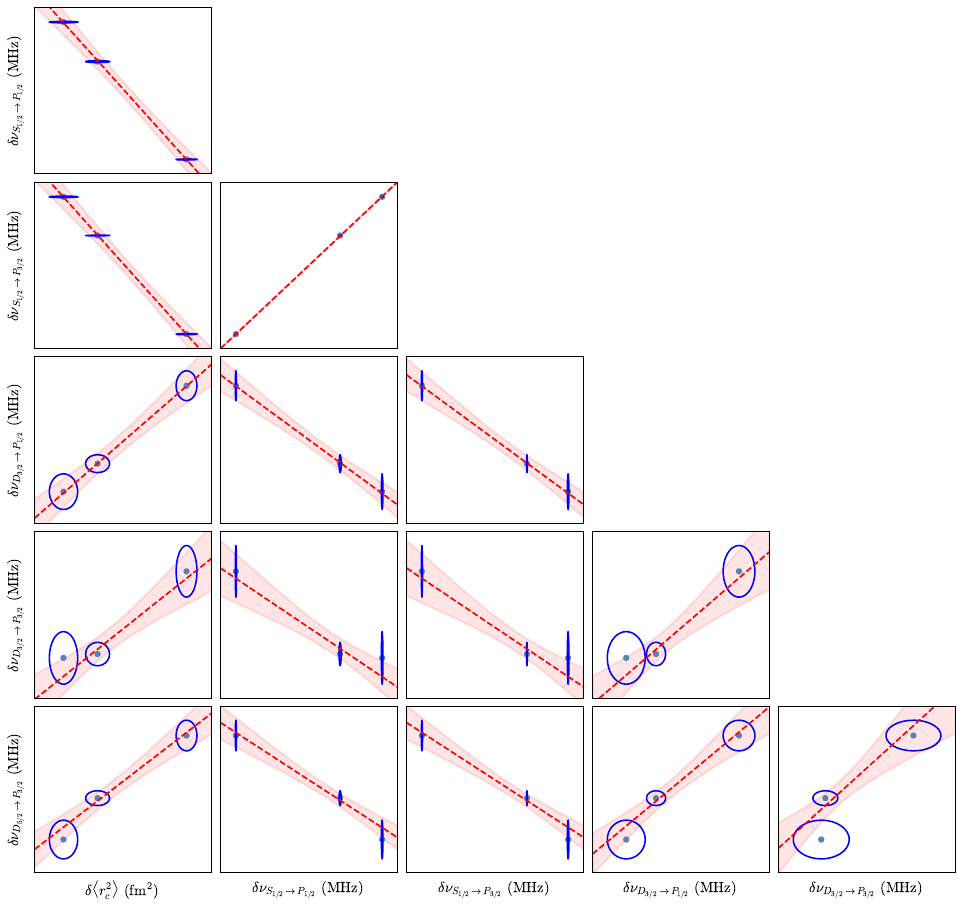}
    \caption{The 2-D projections of the 6-dimensional King plot procedure utilizing the $\chi^2$-fit. The slopes and intercepts of the 2-D projections onto the modified differential charge radii (first column) yield the isotope shift parameters tabulated in Tab.\,\ref{Tab:different_King_plots}. The blue ellipses illustrate the 1$\sigma$-uncertainty of the data points.}
    \label{fig:6d-plot}
\end{figure*}

Table\,\ref{Tab:different_King_plots} presents the results of the experimental and theory-informed King plots together with the corresponding values of $\alpha$ and $K_\alpha$, which were used to calculate the charge radii according to Eq.\,\eqref{eq:charge_radius_extraction}.
\newpage
Additionally, we performed a six-dimensional fit utilizing both Eqs.\,\eqref{eq:isotope_shift} and \eqref{eq:FSR}, thereby simultaneously incorporating the differential charge radii and the optical data from all transitions. This fit was performed both as a $\chi^2$ minimization and with a Monte-Carlo approach using the \texttt{qspec} Python package \cite{muller_qspec_2025}. As both methods yielded indistinguishable results, only the $\chi^2$ fit is included in Tab.\,\ref{Tab:different_King_plots}. While in the two-dimensional case the
correlation between slope and intercept can be removed by introducing a shift of the abscissa by a constant $\alpha$, in the six-dimensional case only an offset vector $\Vec{\alpha}$ can be introduced to reduce, but not eliminate, the parameter correlations. The two-dimensional projections of the resulting $\chi^2$ fit are shown in Fig.\,\ref{fig:6d-plot}.

Although obtained from a single global optimization, the slopes and intercepts listed in Tab.~\ref{Tab:different_King_plots} describe the corresponding two-dimensional projections of the six-dimensional fit. The projections of isotope shift versus differential charge radius therefore directly yield the field- and mass-shift constants $F$ and $K$, respectively. Likewise, the transition-versus-transition King plots have slopes equal to the field-shift ratios introduced in Sec.\,\ref{sec:Method}, Eq.\,\eqref{eq:FSR}, while their intercepts correspond to the appropriate combinations of the field- and mass-shift constants of the two transitions. The results of the six-dimensional fit are nearly identical to those obtained from the individual two-dimensional King plots. However, the fact that all data points are intersected within one standard deviation in Fig.\,\ref{fig:6d-plot} demonstrates the excellent internal consistency of the complete optical data set.

\begin{table*}[t]
\caption{Isotope-shift parameters obtained from the two-dimensional experimental and theory-informed King plots and from the six-dimensional experimental King plot. All values of $F_i$ are given in MHz/fm$^2$, $K_i$ and $K_{\alpha,i}$ in GHz$\cdot$u, and $\alpha$ in $u\cdot\mathrm{fm}^2$. \label{Tab:different_King_plots}}
\centering
\begin{ruledtabular}
\begin{tabular}{c S S S}
{} & \multicolumn{2}{c}{2-D} & {6-D} \\
{} & {Experimental} & {Theory-informed} & {Experimental} \\
\cline{2-3}\cline{4-4}
\hline
F$_{S_{\nicefrac{1}{2}}\rightarrow P_{\nicefrac{1}{2}}}$
    & -997(130) & -1301(10) & -998(130) \\
F$_{S_{\nicefrac{1}{2}}\rightarrow P_{\nicefrac{3}{2}}}$
    & -1001(131) & -1308(5) & -1002(131) \\
F$_{D_{\nicefrac{3}{2}}\rightarrow P_{\nicefrac{1}{2}}}$
    & 221(52) & 265(10) & 221(52) \\
F$_{D_{\nicefrac{3}{2}}\rightarrow P_{\nicefrac{3}{2}}}$
    & 161(59) & 258(10) & 159(59) \\
F$_{D_{\nicefrac{5}{2}}\rightarrow P_{\nicefrac{3}{2}}}$
    & 264(69) & 246(12) & 258(67) \\
\hline
K$_{S_{\nicefrac{1}{2}}\rightarrow P_{\nicefrac{1}{2}}}$
    & 386(40) & 310(31) & 386(46) \\
K$_{S_{\nicefrac{1}{2}}\rightarrow P_{\nicefrac{3}{2}}}$
    & 384(40) & 308(31) & 384(46) \\
K$_{D_{\nicefrac{3}{2}}\rightarrow P_{\nicefrac{1}{2}}}$
    & -1457(17) & -1444(9) & -1457(18) \\
K$_{D_{\nicefrac{3}{2}}\rightarrow P_{\nicefrac{3}{2}}}$
    & -1480(21) & -1450(10) & -1482(20) \\
K$_{D_{\nicefrac{5}{2}}\rightarrow P_{\nicefrac{3}{2}}}$
    & -1437(22) & -1442(11) & -1437(22) \\
\hline
K$_{\alpha,\,S_{\nicefrac{1}{2}}\rightarrow P_{\nicefrac{1}{2}}}$
    & 634(24) & 633(31) & \\
K$_{\alpha,\,S_{\nicefrac{1}{2}}\rightarrow P_{\nicefrac{3}{2}}}$
    & 633(24) & 633(31) & \\
K$_{\alpha,\,D_{\nicefrac{3}{2}}\rightarrow P_{\nicefrac{1}{2}}}$
    & -1519(8) & -1519(9) & \\
K$_{\alpha,\,D_{\nicefrac{3}{2}}\rightarrow P_{\nicefrac{3}{2}}}$
    & -1531(8) & -1532(10) & \\
K$_{\alpha,\,D_{\nicefrac{5}{2}}\rightarrow P_{\nicefrac{3}{2}}}$
    & -1513(10) & -1513(10) & \\
\hline
$\alpha_{S_{\nicefrac{1}{2}}\rightarrow P_{\nicefrac{1}{2}}}$
    & -248.329 & -248.329 & \\
$\alpha_{S_{\nicefrac{1}{2}}\rightarrow P_{\nicefrac{3}{2}}}$
    & -248.497 & -248.497 & \\
$\alpha_{D_{\nicefrac{3}{2}}\rightarrow P_{\nicefrac{1}{2}}}$
    & -283.906 & -283.906 & \\
$\alpha_{D_{\nicefrac{3}{2}}\rightarrow P_{\nicefrac{3}{2}}}$
    & -318.654 & -318.654 & \\
$\alpha_{D_{\nicefrac{5}{2}}\rightarrow P_{\nicefrac{3}{2}}}$
    & -286.381 & -286.381 & \\
\end{tabular}
\end{ruledtabular}
\end{table*}

\clearpage
\bibliography{New_Sr}

@article{Ahmad.1983,
  author  = {Ahmad, S.~A. and Klempt, W. and Neugart, R. and Otten, E.~W. and Wendt, K. and Ekstr{\"o}m, C.},
  title   = {Determination of nuclear spins and moments in a series of radium isotopes},
  journal = {Phys. Lett. B},
  volume  = {133},
  number  = {1-2},
  pages   = {47--52},
  year    = {1983},
  doi     = {10.1016/0370-2693(83)90103-X}
}

@article{Ahmad.1988,
  author  = {Ahmad, S.~A. and Klempt, W. and Neugart, R. and Otten, E.~W. and Reinhard, P.~G. and Ulm, G. and Wendt, K.},
  title = {Mean square charge radii of radium isotopes and octupole deformation in the {$^{220\text{--}228}${Ra}} region}, journal = {Nucl. Phys. A},
  volume  = {483},
  number  = {2},
  pages   = {244--268},
  year    = {1988},
  doi     = {10.1016/0375-9474(88)90534-9}
}

@article{Arnold.1987,
  author  = {Arnold, E. and Bonn, J. and Gegenwart, R. and Neu, W. and Neugart, R. and Otten, E.~W. and Ulm, G. and Wendt, K. and {{ISOLDE Collaboration}}},
  title   = {Nuclear spin and magnetic moment of {$^{11}${Li}}},
  journal = {Phys. Lett. B},
  volume  = {197},
  number  = {3},
  pages   = {311--314},
  year    = {1987},
  doi     = {10.1016/0370-2693(87)90390-X}
}

@article{Arnold.1992,
  author  = {Arnold, E. and Bonn, J. and Klein, A. and Neugart, R. and Neuroth, M. and Otten, E.~W. and Lievens, P. and Reich, H. and Widdra, W. and {{ISOLDE Collaboration}}},
  title   = {Quadrupole moment of {$^{11}${Li}}},
  journal = {Phys. Lett. B},
  volume  = {281},
  pages   = {16--19},
  year    = {1992},
  doi     = {10.1016/0370-2693(92)90266-7}
}

@article{Bouchiat.1985,
  author  = {Bouchiat, M.~A. and Guena, J. and Pottier, L.},
  title   = {Atomic parity violation measurements in the highly forbidden {$6S_{1/2}$--$7S_{1/2}$} caesium transition. {I}. Theoretical analysis, procedure and apparatus},
  journal = {J. Phys. (Paris)},
  volume  = {46},
  number  = {11},
  pages   = {1897--1924},
  year    = {1985},
  doi     = {10.1051/jphys:0198500460110189700}
}

@article{Bouchiat.1986,
  author  = {Bouchiat, M.~A. and Guena, J. and Pottier, L.},
  title   = {Atomic parity violation measurements in the highly forbidden {$6S_{1/2}$--$7S_{1/2}$} caesium transition. {II}. Analysis and control of systematic effects},
  journal = {J. Phys. (Paris)},
  volume  = {47},
  number  = {7},
  pages   = {1175--1202},
  year    = {1986},
  doi     = {10.1051/jphys:019860047070117500}
}

@article{Bouchiat.1997,
  author  = {Bouchiat, M.~A. and Bouchiat, C.},
  title   = {Parity violation in atoms},
  journal = {Rep. Prog. Phys.},
  volume  = {60},
  number  = {11},
  pages   = {1351--1396},
  year    = {1997},
  doi     = {10.1088/0034-4885/60/11/004}
}

@article{Buchinger.1985,
  author  = {Buchinger, F. and Corriveau, R. and Ramsay, E.~B. and Berdichevsky, D. and Sprung, D.~W.~L.},
  title   = {Influence of the {$N=50$} shell closure on mean square charge radii of {Sr}},
  journal = {Phys. Rev. C},
  volume  = {32},
  number  = {6},
  pages   = {2058},
  year    = {1985},
  doi     = {10.1103/PhysRevC.32.2058}
}

@article{Buchinger.1990,
  author  = {Buchinger, F. and Ramsay, E.~B. and Arnold, E. and Neu, W. and Neugart, R. and Wendt, K. and Silverans, R.~E. and Lievens, P. and Vermeeren, L. and Berdichevsky, D. and Fleming, R. and Sprung, D.~W.~L. and Ulm, G.},
  title   = {Systematics of nuclear ground state properties in {$^{78\text{--}100}${Sr}} by laser spectroscopy},
  journal = {Phys. Rev. C},
  volume  = {41},
  pages   = {2883--2897},
  year    = {1990},
  doi     = {10.1103/PhysRevC.41.2883}
}

@article{Campbell.2016,
  author  = {Campbell, P. and Moore, I.~D. and Pearson, M.~R.},
  title   = {Laser spectroscopy for nuclear structure physics},
  journal = {Prog. Part. Nucl. Phys.},
  volume  = {86},
  pages   = {127--180},
  year    = {2016},
  doi     = {10.1016/j.ppnp.2015.09.003}
}

@article{Coc.1985,
  author  = {Coc, A. and Thibault, C. and Touchard, F. and Duong, H.~T. and Juncar, P. and Liberman, S. and Pinard, J. and Lerme, J. and Vialle, J.~L. and Buettgenbach, S. and Mueller, A.~C. and Pesnelle, A.},
  title   = {Hyperfine structures and isotope shifts of {$^{207\text{--}213,220\text{--}228}${Fr}}; possible evidence of octupolar deformation},
  journal = {Phys. Lett. B},
  volume  = {163},
  pages   = {66--70},
  year    = {1985},
  doi     = {10.1016/0370-2693(85)90193-5}
}

@article{Collister.2014,
  author  = {Collister, R. and Gwinner, G. and Tandecki, M. and Behr, J.~A. and Pearson, M.~R. and Zhang, J. and Orozco, L.~A. and Aubin, S. and Gomez, E. and {{FrPNC Collaboration}}},
  title   = {Isotope shifts in francium isotopes {$^{206\text{--}213}${Fr}} and {$^{221}${Fr}}},
  journal = {Phys. Rev. A},
  volume  = {90},
  number  = {5},
  pages   = {052502},
  year    = {2014},
  doi     = {10.1103/PhysRevA.90.052502}
}

@article{DeVries.1987,
  author  = {De Vries, H. and De Jager, C.~W. and De Vries, C.},
  title   = {Nuclear charge-density-distribution parameters from elastic electron scattering},
  journal = {At. Data Nucl. Data Tables},
  volume  = {36},
  number  = {3},
  pages   = {495--536},
  year    = {1987},
  doi     = {10.1016/0092-640X(87)90013-1}
}

@article{Ewald.2004,
  author  = {Ewald, G. and N{\"o}rtersh{\"a}user, W. and Dax, A. and G{\"o}tte, S. and Kirchner, R. and Kluge, H.-J. and K{\"u}hl, T. and Sanchez, R. and Wojtaszek, A. and Bushaw, B.~A. and Drake, G.~W.~F. and Yan, Z.-C. and Zimmermann, C.},
  title   = {Nuclear charge radii of {$^{8,9}${Li}} determined by laser spectroscopy},
  journal = {Phys. Rev. Lett.},
  volume  = {93},
  pages   = {113002},
  year    = {2004},
  doi     = {10.1103/PhysRevLett.93.113002}
}

@book{Fricke.2004,
  author    = {Fricke, G. and Heilig, K.},
  title     = {Nuclear charge radii},
  series    = {Elementary Particles, Nuclei and Atoms},
  volume    = {20},
  publisher = {Springer},
  address   = {Berlin, Heidelberg},
  year      = {2004},
  isbn      = {978-3-540-42829-9}
}

@article{GarciaRuiz.2016,
  author  = {{Garcia Ruiz}, R.~F. and Bissell, M.~L. and Blaum, K. and Ekstr{\"o}m, A. and Frommgen, N. and Hagen, G. and Hammen, M. and Hebeler, K. and Holt, J.~D. and Jansen, G.~R. and Kowalska, M. and Kreim, K. and Nazarewicz, W. and Neugart, R. and Neyens, G. and N{\"o}rtersh{\"a}user, W. and Papenbrock, T. and Papuga, J. and Schwenk, A. and Simonis, J. and Wendt, K.~A. and Yordanov, D.~T.},
  title   = {Unexpectedly large charge radii of neutron-rich calcium isotopes},
  journal = {Nat. Phys.},
  volume  = {12},
  number  = {6},
  pages   = {594--598},
  year    = {2016},
  doi     = {10.1038/nphys3645}
}

@article{Geithner.1999,
  author  = {Geithner, W. and Kappertz, S. and Keim, M. and Lievens, P. and Neugart, R. and Vermeeren, L. and Wilbert, S. and Fedoseyev, V.~N. and K{\"o}ster, U. and Mishin, V.~I. and Sebastian, V. and {{ISOLDE Collaboration}}},
  title   = {Measurement of the magnetic moment of the one-neutron halo nucleus {$^{11}${Be}}},
  journal = {Phys. Rev. Lett.},
  volume  = {83},
  number  = {19},
  pages   = {3792},
  year    = {1999},
  doi     = {10.1103/PhysRevLett.83.3792}
}

@article{Hammen.2018,
  author  = {Hammen, M. and N{\"o}rtersh{\"a}user, W. and Balabanski, D.~L. and Bissell, M.~L. and Blaum, K. and Budin{\v{c}}evi{\'c}, I. and Cheal, B. and Flanagan, K.~T. and Fr{\"o}mmgen, N. and Georgiev, G. and Geppert, Ch. and Kowalska, M. and Kreim, K. and Krieger, A. and Nazarewicz, W. and Neugart, R. and Neyens, G. and Papuga, J. and Reinhard, P.~G. and Rajabali, M.~M. and Schmidt, S. and Yordanov, D.~T.},
  title   = {From calcium to cadmium: testing the pairing functional through charge radii measurements of {$^{100\text{--}130}${Cd}}},
  journal = {Phys. Rev. Lett.},
  volume  = {121},
  number  = {10},
  pages   = {102501},
  year    = {2018},
  doi     = {10.1103/PhysRevLett.121.102501}
}

@article{Huber.1978,
  author  = {Huber, G. and Touchard, F. and B{\"u}ttgenbach, S. and Thibault, C. and Klapisch, R. and Duong, H.~T. and Liberman, S. and Pinard, J. and Vialle, J.~L. and Juncar, P. and Jacquinot, P.},
  title   = {Spins, magnetic moments, and isotope shifts of {$^{21\text{--}31}${Na}} by high-resolution laser spectroscopy of the atomic {$D_1$} line},
  journal = {Phys. Rev. C},
  volume  = {18},
  number  = {5},
  pages   = {2342--2354},
  year    = {1978},
  doi     = {10.1103/PhysRevC.18.2342}
}

@article{Imgram.2019,
  author  = {Imgram, P. and K{\"o}nig, K. and Kr{\"a}mer, J. and Ratajczyk, T. and M{\"u}ller, R.~A. and Surzhykov, A. and N{\"o}rtersh{\"a}user, W.},
  title   = {Collinear laser spectroscopy at ion-trap accuracy: transition frequencies and isotope shifts in the {$6s\,^{2}S_{1/2}$--$6p\,^{2}P_{1/2,3/2}$} transitions in {Ba}$^{+}$},
  journal = {Phys. Rev. A},
  volume  = {99},
  number  = {1},
  pages   = {012511},
  year    = {2019},
  doi     = {10.1103/PhysRevA.99.012511}
}

@article{Imgram.2023c,
  author  = {Imgram, P. and K{\"o}nig, K. and Maa{\ss}, B. and M{\"u}ller, P. and N{\"o}rtersh{\"a}user, W.},
  title   = {Collinear laser spectroscopy of highly charged ions produced with an electron-beam ion source},
  journal = {Phys. Rev. A},
  volume  = {108},
  number  = {6},
  pages   = {062809},
  year    = {2023},
  doi     = {10.1103/PhysRevA.108.062809}
}

@article{Kaufman.1976,
  author  = {Kaufman, S.~L.},
  title   = {High-resolution laser spectroscopy in fast beams},
  journal = {Opt. Commun.},
  volume  = {17},
  number  = {3},
  pages   = {309--312},
  year    = {1976},
  doi     = {10.1016/0030-4018(76)90267-4}
}

@article{Konig.2020b,
  author  = {K{\"o}nig, K. and Kr{\"a}mer, J. and Geppert, C. and Imgram, P. and Maa{\ss}, B. and Ratajczyk, T. and N{\"o}rtersh{\"a}user, W.},
  title   = {A new collinear apparatus for laser spectroscopy and applied science ({COALA})},
  journal = {Rev. Sci. Instrum.},
  volume  = {91},
  number  = {8},
  pages   = {081301},
  year    = {2020},
  doi     = {10.1063/5.0010903}
}

@article{Konig.2021b,
  author  = {K{\"o}nig, K. and Sommer, F. and Lantis, J. and Minamisono, K. and N{\"o}rtersh{\"a}user, W. and Pineda, S. and Powel, R.},
  title   = {Isotope-shift measurements and King-fit analysis in nickel isotopes},
  journal = {Phys. Rev. C},
  volume  = {103},
  number  = {5},
  pages   = {054305},
  year    = {2021},
  doi     = {10.1103/PhysRevC.103.054305}
}

@article{Konig.2024,
  author  = {K{\"o}nig, K. and K{\"o}hler, F. and Palmes, J. and Badura, H. and Dockery, A. and Minamisono, K. and Meisner, J. and M{\"u}ller, P. and N{\"o}rtersh{\"a}user, W. and Passon, S.},
  title   = {High voltage determination and stabilization for collinear laser spectroscopy applications},
  journal = {Rev. Sci. Instrum.},
  volume  = {95},
  pages   = {083307},
  year    = {2024},
  doi     = {10.1063/5.0218649}
}

@article{Koszorus.2021,
  author  = {Koszor{\'u}s, {\'A}. and Yang, X.~F. and Jiang, W.~G. and Novario, S.~J. and Bai, S.~W. and Billowes, J. and Binnersley, C.~L. and Bissell, M.~L. and Cocolios, T.~E. and Cooper, B.~S. and {de Groote}, R.~P. and Ekstr{\"o}m, A. and Flanagan, K.~T. and Forss{\'e}n, C. and Franchoo, S. and {Garcia Ruiz}, R.~F. and Gustafsson, F.~P. and Hagen, G. and Jansen, G.~R. and Kanellakopoulos, A. and Kortelainen, M. and Nazarewicz, W. and Neyens, G. and Papenbrock, T. and Reinhard, P.-G. and Ricketts, C.~M. and Sahoo, B.~K. and Vernon, A.~R. and Wilkins, S.~G.},
  title   = {Charge radii of exotic potassium isotopes challenge nuclear theory and the magic character of {$N=32$}},
  journal = {Nat. Phys.},
  volume  = {17},
  pages   = {439--443},
  year    = {2021},
  doi     = {10.1038/s41567-020-01136-5}
}

@article{Kreim.2014,
  author  = {Kreim, K. and Bissell, M.~L. and Papuga, J. and Blaum, K. and de Rydt, M. and {Garcia Ruiz}, R.~F. and Goriely, S. and Heylen, H. and Kowalska, M. and Neugart, R. and Neyens, G. and N{\"o}rtersh{\"a}user, W. and Rajabali, M.~M. and {S{\'a}nchez Alarc{\'o}n}, R. and Stroke, H.~H. and Yordanov, D.~T.},
  title   = {Nuclear charge radii of potassium isotopes beyond {$N=28$}},
  journal = {Phys. Lett. B},
  volume  = {731},
  pages   = {97--102},
  year    = {2014},
  doi     = {10.1016/j.physletb.2014.02.012}
}

@article{Kretzschmar.2004,
  author  = {Kretzschmar, M. and G{\"o}tte, S. and Ewald, G. and Knaak, K.~M. and Wendt, K.~D.~A. and Kluge, H.~J.},
  title   = {Influence of the thermal motion on the line shape and position of resonances in collinear fast beam laser spectroscopy},
  journal = {Appl. Phys. B},
  volume  = {79},
  number  = {5},
  pages   = {623--627},
  year    = {2004},
  doi     = {10.1007/s00340-004-1608-1}
}

@article{Krieger.2012,
  author  = {Krieger, A. and Blaum, K. and Bissell, M.~L. and Fr{\"o}mmgen, N. and Geppert, Ch. and Hammen, M. and Kreim, K. and Kowalska, M. and Kr{\"a}mer, J. and Neff, T. and Neugart, R. and Neyens, G. and N{\"o}rtersh{\"a}user, W. and Novotny, Ch. and Sanchez, R. and Yordanov, D.~T.},
  title   = {Nuclear charge radius of {$^{12}${Be}}},
  journal = {Phys. Rev. Lett.},
  volume  = {108},
  number  = {14},
  pages   = {142501},
  year    = {2012},
  doi     = {10.1103/PhysRevLett.108.142501}
}

@article{Lynch.2014,
  author  = {Lynch, K.~M. and Billowes, J. and Bissell, M.~L. and Budin{\v{c}}evi{\'c}, I. and Cocolios, T.~E. and {de Groote}, R.~P. and {de Schepper}, S. and Fedosseev, V.~N. and Flanagan, K.~T. and Franchoo, S. and {Garcia Ruiz}, R.~F. and Heylen, H. and Marsh, B.~A. and Neyens, G. and Procter, T.~J. and Rossel, R.~E. and Rothe, S. and Strashnov, I. and Stroke, H.~H. and Wendt, K.~D.~A.},
  title   = {Decay-assisted laser spectroscopy of neutron-deficient francium},
  journal = {Phys. Rev. X},
  volume  = {4},
  pages   = {011055},
  year    = {2014},
  doi     = {10.1103/PhysRevX.4.011055}
}

@article{Mane.2011c,
  author  = {Man{\'e}, E. and Voss, A. and Behr, J.~A. and Billowes, J. and Brunner, T. and Buchinger, F. and Crawford, J.~E. and Dilling, J. and Ettenauer, S. and Levy, C.~D.~P. and Shelbaya, O. and Pearson, M.~R.},
  title   = {First experimental determination of the charge radius of {$^{74}${Rb}} and its application in tests of the unitarity of the Cabibbo--Kobayashi--Maskawa matrix},
  journal = {Phys. Rev. Lett.},
  volume  = {107},
  number  = {21},
  pages   = {212502},
  year    = {2011},
  doi     = {10.1103/PhysRevLett.107.212502}
}

@article{MartenssonPendrill.1992,
  author  = {M{\aa}rtensson-Pendrill, A.-M. and Ynnerman, A. and Warston, H. and Vermeeren, L. and Silverans, R.~E. and Klein, A. and Neugart, R. and Schulz, C. and Lievens, P.},
  title   = {Isotope shifts and nuclear charge radii in singly ionized {$^{40\text{--}48}${Ca}}},
  journal = {Phys. Rev. A},
  volume  = {45},
  pages   = {4675--4681},
  year    = {1992},
  doi     = {10.1103/PhysRevA.45.4675}
}

@article{Miller.2019,
  author  = {Miller, A.~J. and Minamisono, K. and Klose, A. and Garand, D. and Kujawa, C. and Lantis, J.~D. and Liu, Y. and Maa{\ss}, B. and Mantica, P.~F. and Nazarewicz, W. and N{\"o}rtersh{\"a}user, W. and Pineda, S.~V. and Reinhard, P.~G. and Rossi, D.~M. and Sommer, F. and Sumithrarachchi, C. and Teigelh{\"o}fer, A. and Watkins, J.},
  title   = {Proton superfluidity and charge radii in proton-rich calcium isotopes},
  journal = {Nat. Phys.},
  volume  = {15},
  number  = {5},
  pages   = {432--436},
  year    = {2019},
  doi     = {10.1038/s41567-019-0416-9}
}

@article{Mueller.1983,
  author  = {Mueller, A.~C. and Buchinger, F. and Klempt, W. and Otten, E.~W. and Neugart, R. and Ekstr{\"o}m, C. and Heinemeier, J.},
  title   = {Spins, moments and charge radii of barium isotopes in the range {$^{122\text{--}146}${Ba}} determined by collinear fast-beam laser spectroscopy},
  journal = {Nucl. Phys. A},
  volume  = {403},
  number  = {2},
  pages   = {234--262},
  year    = {1983},
  doi     = {10.1016/0375-9474(83)90226-9}
}

@article{Muller.2020,
  author  = {M{\"u}ller, P. and K{\"o}nig, K. and Imgram, P. and Kr{\"a}mer, J. and N{\"o}rtersh{\"a}user, W.},
  title   = {Collinear laser spectroscopy of {Ca}$^{+}$: solving the field-shift puzzle of the {$4s\,^{2}S_{1/2}$--$4p\,^{2}P_{1/2,3/2}$} transitions},
  journal = {Phys. Rev. Research},
  volume  = {2},
  number  = {4},
  pages   = {043351},
  year    = {2020},
  doi     = {10.1103/PhysRevResearch.2.043351}
}

@article{Neyens.2005,
  author  = {Neyens, G. and Kowalska, M. and Yordanov, D. and Blaum, K. and Himpe, P. and Lievens, P. and Mallion, S. and Neugart, R. and Vermeulen, N. and Utsuno, Y. and Otsuka, T.},
  title   = {Measurement of the spin and magnetic moment of {$^{31}${Mg}}: evidence for a strongly deformed intruder ground state},
  journal = {Phys. Rev. Lett.},
  volume  = {94},
  number  = {2},
  pages   = {022501},
  year    = {2005},
  doi     = {10.1103/PhysRevLett.94.022501}
}

@article{Nortershauser.2009,
  author  = {N{\"o}rtersh{\"a}user, W. and Tiedemann, D. and Zakova, M. and Andjelkovic, Z. and Blaum, K. and Bissell, M.~L. and Cazan, R. and Drake, G.~W.~F. and Geppert, C. and Kowalska, M. and Kr{\"a}mer, J. and Krieger, A. and Neugart, R. and Sanchez, R. and Schmidt-Kaler, F. and Yan, Z.-C. and Yordanov, D.~T. and Zimmermann, C.},
  title   = {Nuclear charge radii of {$^{7,9,10}${Be}} and the one-neutron halo nucleus {$^{11}${Be}}},
  journal = {Phys. Rev. Lett.},
  volume  = {102},
  number  = {6},
  pages   = {062503},
  year    = {2009},
  doi     = {10.1103/PhysRevLett.102.062503}
}

@article{Papuga.2013,
  author  = {Papuga, J. and Bissell, M.~L. and Kreim, K. and Blaum, K. and Brown, B.~A. and de Rydt, M. and {Garcia Ruiz}, R.~F. and Heylen, H. and Kowalska, M. and Neugart, R. and Neyens, G. and N{\"o}rtersh{\"a}user, W. and Otsuka, T. and Rajabali, M.~M. and S{\'a}nchez, R. and Utsuno, Y. and Yordanov, D.~T.},
  title   = {Spins and magnetic moments of {$^{49,51}${K}}: establishing the {$1/2^{+}$} and {$3/2^{+}$} level ordering beyond {$N=28$}},
  journal = {Phys. Rev. Lett.},
  volume  = {110},
  number  = {17},
  pages   = {172503},
  year    = {2013},
  doi     = {10.1103/PhysRevLett.110.172503}
}

@article{Schinzler.1978,
  author  = {Schinzler, B. and Klempt, W. and Kaufman, S.~L. and Lochmann, H. and Moruzzi, G. and Neugart, R. and Otten, E.~W. and Bonn, J. and {von Reisky}, L. and Spath, K.~P.~C.},
  title   = {Collinear laser spectroscopy of neutron-rich {Cs} isotopes at an on-line mass separator},
  journal = {Phys. Lett. B},
  volume  = {79},
  number  = {3},
  pages   = {209--212},
  year    = {1978},
  doi     = {10.1016/0370-2693(78)90224-1}
}

@article{Shi.2016,
  author  = {Shi, C. and Gebert, F. and Gorges, C. and Kaufmann, S. and N{\"o}rtersh{\"a}user, W. and Sahoo, B.~K. and Surzhykov, A. and Yerokhin, V.~A. and Berengut, J.~C. and Wolf, F. and Heip, J.~C. and Schmidt, P.~O.},
  title   = {Unexpectedly large difference of the electron density at the nucleus in the {$4p\,^{2}P_{1/2,3/2}$} fine-structure doublet of {Ca}$^{+}$},
  journal = {Appl. Phys. B},
  volume  = {123},
  pages   = {2},
  year    = {2016},
  doi     = {10.1007/s00340-016-6572-z}
}

@article{Takamine.2014,
  author  = {Takamine, A. and Wada, M. and Okada, K. and Sonoda, T. and Schury, P. and Nakamura, T. and Kanai, Y. and Kubo, T. and Katayama, I. and Ohtani, S. and Wollnik, H. and Schuessler, H.~A.},
  title   = {Hyperfine structure constant of the neutron halo nucleus {$^{11}${Be}}$^{+}$},
  journal = {Phys. Rev. Lett.},
  volume  = {112},
  number  = {16},
  pages   = {162502},
  year    = {2014},
  doi     = {10.1103/PhysRevLett.112.162502}
}

@article{Thibault.1981,
  author  = {Thibault, C. and Touchard, F. and B{\"u}ttgenbach, S. and Klapisch, R. and {de Saint Simon}, M. and Duong, H.~T. and Jacquinot, P. and Juncar, P. and Liberman, S. and Pillet, P. and Pinard, J. and Vialle, J.~L. and Pesnelle, A. and Huber, G.},
  title   = {Hyperfine structure and isotope shift of the {$D_2$} line of {$^{76\text{--}98}${Rb}} and some of their isomers},
  journal = {Phys. Rev. C},
  volume  = {23},
  number  = {6},
  pages   = {2720},
  year    = {1981},
  doi     = {10.1103/PhysRevC.23.2720}
}

@article{Wendt.1984,
  author  = {Wendt, K. and Ahmad, S.~A. and Buchinger, F. and Mueller, A.~C. and Neugart, R. and Otten, E.~W.},
  title   = {Relativistic {$J$}-dependence of the isotope shift in the {$6s$--$6p$} doublet of {Ba}~{II}},
  journal = {Z. Phys. A},
  volume  = {318},
  number  = {2},
  pages   = {125--129},
  year    = {1984},
  doi     = {10.1007/BF01413460}
}

@article{Wendt.1988,
  author  = {Wendt, K. and Ahmad, S.~A. and Ekstr{\"o}m, C. and Klempt, W. and Neugart, R. and Otten, E.~W.},
  title   = {Hyperfine structure and isotope shift of the neutron-rich barium isotopes {$^{139\text{--}146}${Ba}} and {$^{148}${Ba}}},
  journal = {Z. Phys. A},
  volume  = {329},
  number  = {4},
  pages   = {407--411},
  year    = {1988},
  doi     = {10.1007/BF01294345}
}

@article{Wood.1997,
  author  = {Wood, C.~S. and Bennett, S.~C. and Cho, D. and Masterson, B.~P. and Roberts, J.~L. and Tanner, C.~E. and Wieman, C.~E.},
  title   = {Measurement of parity nonconservation and an anapole moment in cesium},
  journal = {Science},
  volume  = {275},
  number  = {5307},
  pages   = {1759--1763},
  year    = {1997},
  doi     = {10.1126/science.275.5307.1759}
}

@article{Yang.2023,
  author  = {Yang, X.~F. and Wang, S.~J. and Wilkins, S.~G. and {Garcia Ruiz}, R.~F.},
  title   = {Laser spectroscopy for the study of exotic nuclei},
  journal = {Prog. Part. Nucl. Phys.},
  volume  = {129},
  pages   = {104005},
  year    = {2023},
  doi     = {10.1016/j.ppnp.2022.104005}
}

@phdthesis{Yordanov.2007,
  author  = {Yordanov, D.},
  title   = {From {$^{27}${Mg}} to {$^{33}${Mg}}: transition to the ``island of inversion''},
  school  = {Katholieke Universiteit Leuven},
  address = {Leuven},
  year    = {2007}
}

@article{Yordanov.2012,
  author  = {Yordanov, D.~T. and Bissell, M.~L. and Blaum, K. and de Rydt, M. and Geppert, Ch. and Kowalska, M. and Kr{\"a}mer, J. and Kreim, K. and Krieger, A. and Lievens, P. and Neff, T. and Neugart, R. and Neyens, G. and N{\"o}rtersh{\"a}user, W. and Sanchez, R. and Vingerhoets, P.},
  title   = {Nuclear charge radii of {$^{21\text{--}32}${Mg}}},
  journal = {Phys. Rev. Lett.},
  volume  = {108},
  number  = {4},
  pages   = {042504},
  year    = {2012},
  doi     = {10.1103/PhysRevLett.108.042504}
}

@article{York.2004,
  author  = {York, D. and Evensen, N.~M. and Mart{\'\i}nez, M.~L. and {de Basabe Delgado}, J.},
  title   = {Unified equations for the slope, intercept, and standard errors of the best straight line},
  journal = {Am. J. Phys.},
  volume  = {72},
  number  = {3},
  pages   = {367--375},
  year    = {2004},
  doi     = {10.1119/1.1632486}
}

@article{virtanen2020scipy,
  author  = {Virtanen, P. and Gommers, R. and Oliphant, T.~E. and Haberland, M. and Reddy, T. and Cournapeau, D. and Burovski, E. and Peterson, P. and Weckesser, W. and Bright, J. and {van der Walt}, S.~J. and Brett, M. and Wilson, J. and Millman, K.~J. and Mayorov, N. and Nelson, A.~R.~J. and Jones, E. and Kern, R. and Larson, E. and Carey, C.~J. and Polat, {\.I}. and Feng, Y. and Moore, E.~W. and VanderPlas, J. and Laxalde, D. and Perktold, J. and Cimrman, R. and Henriksen, I. and Quintero, E.~A. and Harris, C.~R. and Archibald, A.~M. and Ribeiro, A.~H. and Pedregosa, F. and {van Mulbregt}, P.},
  title   = {{SciPy} 1.0: fundamental algorithms for scientific computing in {Python}},
  journal = {Nat. Methods},
  volume  = {17},
  number  = {3},
  pages   = {261--272},
  year    = {2020},
  doi     = {10.1038/s41592-019-0686-2}
}

@article{barwood_observation_2003,
  author  = {Barwood, G.~P. and Gao, K. and Gill, P. and Huang, G. and Klein, H.~A.},
  title   = {Observation of the hyperfine structure of the {$2\,^{2}S_{1/2}$--$2\,^{2}D_{5/2}$} transition in {$^{87}${Sr}}$^{+}$},
  journal = {Phys. Rev. A},
  volume  = {67},
  number  = {1},
  pages   = {013402},
  year    = {2003},
  doi     = {10.1103/PhysRevA.67.013402}
}

@article{yu_calculation_2004,
  author  = {Yu, K.-Z. and Wu, L.-J. and Gou, B.-C. and Shi, T.-Y.},
  title   = {Calculation of the hyperfine structure constants in {$^{43}${Ca}}$^{+}$ and {$^{87}${Sr}}$^{+}$},
  journal = {Phys. Rev. A},
  volume  = {70},
  number  = {1},
  pages   = {012506},
  year    = {2004},
  doi     = {10.1103/PhysRevA.70.012506}
}

@article{sunaoshi_precision_1993,
  author  = {Sunaoshi, H. and Fukashiro, Y. and Furukawa, M. and Yamauchi, M. and Hayashibe, S. and Shinozuka, T. and Fujioka, M. and Satoh, I. and Wada, M. and Matsuki, S.},
  title   = {A precision measurement of the hyperfine structure of {$^{87}${Sr}}$^{+}$},
  journal = {Hyperfine Interact.},
  volume  = {78},
  number  = {1-4},
  pages   = {241--245},
  year    = {1993},
  doi     = {10.1007/BF00568145}
}

@article{likforman_precision_2016,
  author  = {Likforman, J.-P. and Tugay{\'e}, V. and Guibal, S. and Guidoni, L.},
  title   = {Precision measurement of the branching fractions of the {$5p\,^{2}P_{1/2}$} state in {$^{88}${Sr}}$^{+}$ with a single ion in a microfabricated surface trap},
  journal = {Phys. Rev. A},
  volume  = {93},
  number  = {5},
  pages   = {052507},
  year    = {2016},
  doi     = {10.1103/PhysRevA.93.052507}
}

@article{zhang_iterative_2016,
  author  = {Zhang, H. and Gutierrez, M. and Low, G.~H. and Rines, R. and Stuart, J. and Wu, T. and Chuang, I.},
  title   = {Iterative precision measurement of branching ratios applied to {$5P$} states in {$^{88}${Sr}}$^{+}$},
  journal = {New J. Phys.},
  volume  = {18},
  number  = {12},
  pages   = {123021},
  year    = {2016},
  doi     = {10.1088/1367-2630/aa511d}
}

@article{pinnington_studies_1995,
  author  = {Pinnington, E.~H. and Berends, R.~W. and Lumsden, M.},
  title   = {Studies of laser-induced fluorescence in fast beams of {Sr}$^{+}$ and {Ba}$^{+}$ ions},
  journal = {J. Phys. B: At. Mol. Opt. Phys.},
  volume  = {28},
  pages   = {2095--2103},
  year    = {1995},
  doi     = {10.1088/0953-4075/28/11/009}
}

@article{letchumanan_lifetime_2005,
  author  = {Letchumanan, V. and Wilson, M.~A. and Gill, P. and Sinclair, A.~G.},
  title   = {Lifetime measurement of the metastable {$4d\,^{2}D_{5/2}$} state in {$^{88}${Sr}}$^{+}$ using a single trapped ion},
  journal = {Phys. Rev. A},
  volume  = {72},
  number  = {1},
  pages   = {012509},
  year    = {2005},
  doi     = {10.1103/PhysRevA.72.012509}
}

@article{biemont_lifetimes_2000,
  author  = {Bi{\'e}mont, E. and Lidberg, J. and Mannervik, S. and Norlin, L.-O. and Royen, P. and Schmitt, A. and Shi, W. and Tordoir, X.},
  title   = {Lifetimes of metastable states in {Sr}~{II}},
  journal = {Eur. Phys. J. D},
  volume  = {11},
  number  = {3},
  pages   = {355--365},
  year    = {2000},
  doi     = {10.1007/s100530070063}
}

@article{margolis_absolute_2003,
  author  = {Margolis, H.~S. and Huang, G. and Barwood, G.~P. and Lea, S.~N. and Klein, H.~A. and Rowley, W.~R.~C. and Gill, P. and Windeler, R.~S.},
  title   = {Absolute frequency measurement of the 674-nm {$^{88}${Sr}}$^{+}$ clock transition using a femtosecond optical frequency comb},
  journal = {Phys. Rev. A},
  volume  = {67},
  number  = {3},
  pages   = {032501},
  year    = {2003},
  doi     = {10.1103/PhysRevA.67.032501}
}

@article{Hosier2025,
  author  = {Hosier, A. and Dipti and Blundell, S.~A. and Lapierre, A. and Silwal, R. and Gwinner, G. and Tan, J.~N. and Naing, A. and Gillaspy, J.~D. and Yang, Y. and Szypryt, P. and O'Neil, G. and Staiger, H. and Dreiling, J.~M. and Villari, A.~C.~C. and Angeli, I. and Ralchenko, Yu. and Takacs, E.},
  title   = {Determination of nuclear charge radius by extreme-ultraviolet spectroscopy of {Na}-like ions},
  journal = {Phys. Rev. Research},
  volume  = {7},
  pages   = {L012024},
  year    = {2025},
  doi     = {10.1103/PhysRevResearch.7.L012024}
}

@article{Staiger2025,
  author  = {Staiger, H. and Mondeel, G. and Blundell, S.~A. and Dipti and O'Neil, G. and Silwal, R. and Lapierre, A. and Gwinner, G. and Tan, J.~N. and Gillaspy, J.~D. and Ralchenko, Yu. and Takacs, E.},
  title   = {Measurement of {$D$}-line energies in sodiumlike {Ir}},
  journal = {Phys. Rev. A},
  volume  = {112},
  number  = {1},
  pages   = {012807},
  year    = {2025},
  doi     = {10.1103/xf52-d6s6}
}

@article{Silwal2025,
  author  = {Silwal, R. and Takacs, E. and Wang, Y. and Kwiatkowski, A.~A. and Blundell, S.~A. and Staiger, H. and Dipti and Lapierre, A. and Cardona, J.~D. and Maldonado Mill{\'a}n, F. and O'Neil, G. and Hosier, A. and Gillaspy, J.~D. and Ralchenko, Yu. and Gwinner, G.},
  title   = {Highly charged ion approach to measure nuclear charge radii of {Fr}, {Ra}, and {Rn} isotopes for precision measurements},
  journal = {Nucl. Instrum. Methods Phys. Res. A},
  volume  = {1082},
  pages   = {170947},
  year    = {2025},
  doi     = {10.1016/j.nima.2025.170947}
}

@article{lybarger_precision_2011,
  author  = {Lybarger, W.~E. and Berengut, J.~C. and Chiaverini, J.},
  title   = {Precision measurement of the {$5\,^{2}S_{1/2}$--$4\,^{2}D_{5/2}$} quadrupole transition isotope shift between {$^{88}${Sr}}$^{+}$ and {$^{86}${Sr}}$^{+}$},
  journal = {Phys. Rev. A},
  volume  = {83},
  number  = {5},
  pages   = {052509},
  year    = {2011},
  doi     = {10.1103/PhysRevA.83.052509}
}

@article{Ohayon.2025,
  author  = {Ohayon, B.},
  title   = {Critical evaluation of reference charge radii and applications in mirror nuclei},
  journal = {At. Data Nucl. Data Tables},
  volume  = {165},
  pages   = {101732},
  year    = {2025},
  doi     = {10.1016/j.adt.2025.101732}
}

@article{Munro-Laylim.2022,
  author  = {Munro-Laylim, P. and Dzuba, V.~A. and Flambaum, V.~V.},
  title   = {Nuclear polarization and the contributions of relativistic effects to King plot nonlinearity},
  journal = {Phys. Rev. A},
  volume  = {105},
  number  = {4},
  pages   = {042814},
  year    = {2022},
  doi     = {10.1103/PhysRevA.105.042814}
}

@article{Bonn.1979,
  author  = {Bonn, J. and Klempt, W. and Neugart, R. and Otten, E.~W. and Schinzler, B.},
  title   = {Hyperfine structure and isotope shifts of neutron-rich {$^{138\text{--}142}${Cs}}},
  journal = {Z. Phys. A},
  volume  = {289},
  number  = {2},
  pages   = {227--228},
  year    = {1979},
  doi     = {10.1007/BF01435942}
}

@article{dreher_isotopic_1975,
  author  = {Dreher, B.},
  title   = {Isotopic differences in the charge distribution of even molybdenum isotopes from elastic electron scattering},
  journal = {Phys. Rev. Lett.},
  volume  = {35},
  number  = {11},
  pages   = {716--719},
  year    = {1975},
  doi     = {10.1103/PhysRevLett.35.716}
}

@phdthesis{Mallot.1985,
  author  = {Mallot, G.},
  title   = {Elektronenstreuung an Germanium und myonische Krypton-Atome als Beitrag zur Systematik der Kernladungsdichten},
  school  = {Universit{\"a}t Mainz},
  year    = {1985}
}

@techreport{laan_electron_1986,
  author      = {{van der Laan}, J.~B.},
  title       = {Electron scattering off palladium isotopes},
  institution = {University of Amsterdam},
  number      = {INIS-mf--10848},
  year        = {1986}
}

@article{Fricke.1995,
  author  = {Fricke, G. and Bernhardt, C. and Heilig, K. and Schaller, L.~A. and Schellenberg, L. and Shera, E.~B. and {de Jager}, C.~W.},
  title   = {Nuclear ground state charge radii from electromagnetic interactions},
  journal = {At. Data Nucl. Data Tables},
  volume  = {60},
  number  = {2},
  pages   = {177--285},
  year    = {1995},
  doi     = {10.1006/adnd.1995.1007}
}

@article{Toh.2019,
  author  = {Toh, G. and Damitz, A. and Tanner, C.~E. and Johnson, W.~R. and Elliott, D.~S.},
  title   = {Determination of the scalar and vector polarizabilities of the cesium {$6s\,^{2}S_{1/2}$--$7s\,^{2}S_{1/2}$} transition and implications for atomic parity nonconservation},
  journal = {Phys. Rev. Lett.},
  volume  = {123},
  pages   = {073002},
  year    = {2019},
  doi     = {10.1103/PhysRevLett.123.073002}
}

@article{Bennett.1999,
  author  = {Bennett, S.~C. and Wieman, C.~E.},
  title   = {Measurement of the {$6S$--$7S$} transition polarizability in atomic cesium and an improved test of the standard model},
  journal = {Phys. Rev. Lett.},
  volume  = {83},
  pages   = {889},
  year    = {1999},
  doi     = {10.1103/PhysRevLett.83.889}
}

@article{Tah.2023,
  author  = {Tran Tan, H.~B. and Xiao, D. and Derevianko, A.},
  title   = {Reevaluation of Stark-induced transition polarizabilities in cesium},
  journal = {Phys. Rev. A},
  volume  = {108},
  pages   = {022808},
  year    = {2023},
  doi     = {10.1103/PhysRevA.108.022808}
}

@article{Dzuba.2012,
  author  = {Dzuba, V.~A. and Berengut, J.~C. and Flambaum, V.~V. and Roberts, B.},
  title   = {Revisiting parity nonconservation in cesium},
  journal = {Phys. Rev. Lett.},
  volume  = {109},
  pages   = {203003},
  year    = {2012},
  doi     = {10.1103/PhysRevLett.109.203003}
}

@article{Sanamyan.2023,
  author  = {Sanamyan, G. and Roberts, B.~M. and Ginges, J.~S.~M.},
  title   = {Empirical determination of the Bohr--Weisskopf effect in cesium and improved tests of precision atomic theory in searches for new physics},
  journal = {Phys. Rev. Lett.},
  volume  = {130},
  pages   = {053001},
  year    = {2023},
  doi     = {10.1103/PhysRevLett.130.053001}
}

@article{Door.2025,
  author  = {Door, M. and Yeh, C.-H. and Heinz, M. and Kirk, F. and Lyu, C. and Miyagi, T. and Berengut, J.~C. and Biero{\'n}, J. and Blaum, K. and Dreissen, L.~S. and Eliseev, S. and Filianin, P. and Filzinger, M. and Fuchs, E. and F{\"u}rst, H.~A. and Gaigalas, G. and Harman, Z. and Herkenhoff, J. and Huntemann, N. and Keitel, C.~H. and Kromer, K. and Lange, D. and Rischka, A. and Schweiger, C. and Schwenk, A. and Shimizu, N. and Mehlst{\"a}ubler, T.~E.},
  title   = {Probing new bosons and nuclear structure with ytterbium isotope shifts},
  journal = {Phys. Rev. Lett.},
  volume  = {134},
  pages   = {063002},
  year    = {2025},
  doi     = {10.1103/PhysRevLett.134.063002}
}

@article{Steinel.2023,
  author  = {Steinel, M. and Shao, H. and Filzinger, M. and Lipphardt, B. and Brinkmann, M. and Didier, A. and Mehlst{\"a}ubler, T.~E. and Lindvall, T. and Peik, E. and Huntemann, N.},
  title   = {Evaluation of a {$^{88}${Sr}}$^{+}$ optical clock with a direct measurement of the blackbody radiation shift and determination of the clock frequency},
  journal = {Phys. Rev. Lett.},
  volume  = {131},
  pages   = {083002},
  year    = {2023},
  doi     = {10.1103/PhysRevLett.131.083002}
}

@article{Manovitz.2022,
  author  = {Manovitz, T. and Shapira, Y. and Gazit, L. and Akerman, N. and Ozeri, R.},
  title   = {Trapped-ion quantum computer with robust entangling gates and quantum coherent feedback},
  journal = {PRX Quantum},
  volume  = {3},
  pages   = {010347},
  year    = {2022},
  doi     = {10.1103/PRXQuantum.3.010347}
}

@article{Berengut.2018,
  author  = {Berengut, J.~C. and Budker, D. and Delaunay, C. and Flambaum, V.~V. and Frugiuele, C. and Fuchs, E. and Grojean, C. and Harnik, R. and Ozeri, R. and Perez, G. and Soreq, Y.},
  title   = {Probing new long-range interactions by isotope shift spectroscopy},
  journal = {Phys. Rev. Lett.},
  volume  = {120},
  pages   = {091801},
  year    = {2018},
  doi     = {10.1103/PhysRevLett.120.091801}
}

@article{Miyake.2019,
  author  = {Miyake, H. and Pisenti, N.~C. and Elgee, P.~K. and Sitaram, A. and Campbell, G.~K.},
  title   = {Isotope-shift spectroscopy of the {$^{1}S_{0}$--$^{3}P_{1}$} and {$^{1}S_{0}$--$^{3}P_{0}$} transitions in strontium},
  journal = {Phys. Rev. Research},
  volume  = {1},
  pages   = {033113},
  year    = {2019},
  doi     = {10.1103/PhysRevResearch.1.033113}
}

@book{condon1935theory,
  author    = {Condon, E.~U. and Shortley, G.~H.},
  title     = {The theory of atomic spectra},
  publisher = {Cambridge University Press},
  address   = {Cambridge},
  year      = {1935}
}

@article{Jung.2017,
  author  = {Jung, K. and Iwata, Y. and Miyabe, M. and Yamamoto, K. and Yonezu, T. and Wakaida, I. and Hasegawa, S.},
  title   = {Laser cooling and imaging of individual radioactive {$^{90}${Sr}}$^{+}$ ions},
  journal = {Phys. Rev. A},
  volume  = {96},
  pages   = {043424},
  year    = {2017},
  doi     = {10.1103/PhysRevA.96.043424}
}

@article{Jung.2017b,
  author  = {Jung, K. and Yamamoto, K. and Yamamoto, Y. and Miyabe, M. and Wakaida, I. and Hasegawa, S.},
  title   = {All-diode-laser cooling of {Sr}$^{+}$ isotope ions for analytical applications},
  journal = {Jpn. J. Appl. Phys.},
  volume  = {56},
  pages   = {062401},
  year    = {2017},
  doi     = {10.7567/JJAP.56.062401}
}

@article{Dubost.2014,
  author  = {Dubost, B. and Dubessy, R. and Szymanski, B. and Guibal, S. and Likforman, J.-P. and Guidoni, L.},
  title   = {Isotope shifts of natural {Sr}$^{+}$ measured by laser fluorescence in a sympathetically cooled Coulomb crystal},
  journal = {Phys. Rev. A},
  volume  = {89},
  pages   = {032504},
  year    = {2014},
  doi     = {10.1103/PhysRevA.89.032504}
}

@article{Wendt.1997,
  author  = {Wendt, K. and Bhowmick, G.~K. and Bushaw, B.~A. and Herrmann, G. and Kratz, J.~V. and Lantzsch, J. and M{\"u}ller, P. and N{\"o}rtersh{\"a}user, W. and Otten, E.-W. and Schwalbach, R. and Seibert, U.-A. and Trautmann, N. and Waldek, A.},
  title   = {Rapid trace analysis of {$^{89,90}${Sr}} in environmental samples by collinear laser resonance ionization mass spectrometry},
  journal = {Radiochim. Acta},
  volume  = {79},
  pages   = {183--190},
  year    = {1997},
  doi     = {10.1524/ract.1997.79.3.183}
}

@article{Bushaw.2000,
  author  = {Bushaw, B.~A. and N{\"o}rtersh{\"a}user, W.},
  title   = {Resonance ionization spectroscopy of stable strontium isotopes and {$^{90}${Sr}} via {$5s^{2}\,^{1}S_{0}\rightarrow 5s5p\,^{1}P_{1}\rightarrow 5s5d\,^{1}D_{2}\rightarrow 5s11f\,^{1}F_{3}\rightarrow \mathrm{Sr}^{+}$}},
  journal = {Eur. Phys. J. D},
  volume  = {55},
  pages   = {1679--1692},
  year    = {2000},
  doi     = {10.1016/S0584-8547(00)00269-X}
}

@article{Lu.2003,
  author  = {Lu, Z.-T. and Wendt, K.~D.~A.},
  title   = {Laser-based methods for ultrasensitive trace-isotope analyses},
  journal = {Rev. Sci. Instrum.},
  volume  = {74},
  pages   = {1169--1179},
  year    = {2003},
  doi     = {10.1063/1.1535232}
}

@article{Nortershauser.1998b,
  author  = {N{\"o}rtersh{\"a}user, W. and Blaum, K. and Icker, K. and M{\"u}ller, P. and Schmitt, A. and Wendt, K. and Wiche, B.},
  title   = {Isotope shifts and hyperfine structure in transitions in calcium {II}},
  journal = {Eur. Phys. J. D},
  volume  = {2},
  pages   = {33--39},
  year    = {1998},
  doi     = {10.1007/s100530050107}
}

@article{SahooYb+,
  author  = {Sahoo, B.~K.},
  title   = {Precise determination of electric quadrupole moments and isotope shift constants of {Yb}$^{+}$ in pursuance of probing fundamental physics and nuclear radii},
  journal = {Phys. Rev. A},
  volume  = {111},
  number  = {6},
  pages   = {L060801},
  year    = {2025},
  doi     = {10.1103/PhysRevA.111.L060801}
}

@article{SahooAR,
  author  = {Sahoo, B.~K. and Vernon, A.~R. and {Garcia Ruiz}, R.~F. and Binnersley, C.~L. and Billowes, J. and Bissell, M.~L. and Cocolios, T.~E. and Farooq-Smith, G.~J. and Flanagan, K.~T. and Gins, W. and de Groote, R.~P. and Koszor{\'u}s, {\'A}. and Neyens, G. and Lynch, K.~M. and Parnefjord-Gustafsson, F. and Ricketts, C.~M. and Wendt, K.~D.~A. and Wilkins, S.~G. and Yang, X.~F.},
  title   = {Analytic response relativistic coupled-cluster theory: the first application to indium isotope shifts},
  journal = {New J. Phys.},
  volume  = {22},
  number  = {1},
  pages   = {012001},
  year    = {2020},
  doi     = {10.1088/1367-2630/ab66dd}
}

@article{Sahoo1,
  author  = {Sahoo, B.~K. and Ohayon, B.},
  title   = {Benchmarking many-body approaches for the determination of isotope-shift constants: application to {$\mathrm{Li}$, $\mathrm{Be}^{+}$, and $\mathrm{Ar}^{15+}$} isoelectronic systems},
  journal = {Phys. Rev. A},
  volume  = {103},
  number  = {5},
  pages   = {052802},
  year    = {2021},
  doi     = {10.1103/PhysRevA.103.052802}
}

@article{Sahoo2,
  author  = {Sahoo, B.~K. and Blundell, S. and Oleynichenko, A.~V. and {Garcia Ruiz}, R.~F. and Skripnikov, L.~V. and Ohayon, B.},
  title   = {Recent advancements in atomic many-body methods for high-precision studies of isotope shifts},
  journal = {J. Phys. B},
  volume  = {58},
  number  = {4},
  pages   = {042001},
  year    = {2025},
  doi     = {10.1088/1361-6455/adacc1}
}

@article{Stone2025,
  author  = {Stone, N.~J.},
  title   = {Table of nuclear magnetic dipole and electric quadrupole moments},
  journal = {At. Data Nucl. Data Tables},
  volume  = {90},
  number  = {1},
  pages   = {75--176},
  year    = {2005},
  doi     = {10.1016/j.adt.2005.04.001}
}

@article{Sahoo2006,
  author  = {Sahoo, B.~K.},
  title   = {Determination of the nuclear quadrupole moment of {$^{87}${Sr}}},
  journal = {Phys. Rev. A},
  volume  = {73},
  number  = {6},
  pages   = {062501},
  year    = {2006},
  doi     = {10.1103/PhysRevA.73.062501}
}

@article{Chakraborty.2026,
  author  = {Chakraborty, A. and Katyal, V. and Sahoo, B.~K.},
  title   = {Investigating roles of triple excitations for high-precision determination of clock properties of alkaline-earth-metal singly charged ions},
  journal = {Phys. Rev. A},
  volume  = {113},
  pages   = {L011101},
  year    = {2026},
  doi     = {10.1103/zh99-b1m3}
}

@article{borghs_hyperfine_1983,
  author  = {Borghs, G. and De Bisschop, P. and Van Hove, M. and Silverans, R.~E.},
  title   = {Hyperfine interactions in the alkaline-earth {Sr} ions by collinear fast beam laser spectroscopy},
  journal = {Hyperfine Interact.},
  volume  = {15},
  number  = {1},
  pages   = {177--180},
  year    = {1983},
  doi     = {10.1007/BF02159735}
}

@article{wang_ame_2021,
  author  = {Wang, M. and Huang, W.~J. and Kondev, F.~G. and Audi, G. and Naimi, S.},
  title   = {{AME} 2020 atomic mass evaluation (II): tables, graphs and references},
  journal = {Chin. Phys. C},
  volume  = {45},
  number  = {3},
  pages   = {030003},
  year    = {2021},
  doi     = {10.1088/1674-1137/abddaf}
}

@article{ge_high-precision_2024,
  author  = {Ge, Z. and Bai, S. and Eronen, T. and Jokinen, A. and Kankainen, A. and Kujanp{\"a}{\"a}, S. and Moore, I. and Nesterenko, D. and Reponen, M.},
  title   = {High-precision measurement of the atomic mass of {$^{84}${Sr}} and implications to isotope shift studies},
  journal = {Eur. Phys. J. A},
  volume  = {60},
  number  = {7},
  pages   = {147},
  year    = {2024},
  doi     = {10.1140/epja/s10050-024-01359-7}
}

@article{sun_208pb_2025,
  author  = {Sun, Z. and Beyer, K.~A. and Mandrykina, Z.~A. and Valuev, I.~A. and Keitel, C.~H. and Oreshkina, N.~S.},
  title   = {{$^{208}${Pb}} nuclear charge radius revisited: closing the fine-structure-anomaly gap},
  journal = {Phys. Rev. Lett.},
  volume  = {135},
  pages   = {163002},
  year    = {2025},
  doi     = {10.1103/h3xz-xdxr}
}

@article{muller_qspec_2025,
  author  = {M{\"u}ller, P. and N{\"o}rtersh{\"a}user, W.},
  title   = {The {qspec} Python package: a physics toolbox for laser spectroscopy},
  journal = {Comput. Phys. Commun.},
  volume  = {311},
  pages   = {109550},
  year    = {2025},
  doi     = {10.1016/j.cpc.2025.109550}
}

@article{Gorchtein.2026,
  author  = {Gorchtein, M.},
  title   = {Guide to nuclear polarization in muonic atoms},
  journal = {Phys. Rev. C},
  volume  = {113},
  pages   = {L011301},
  year    = {2026},
  doi     = {10.1103/jqm6-c75w}
}

@article{Jian.2023,
  author  = {Jian, B. and Bernard, J. and Gertsvolf, M. and Dub{\'e}, P.},
  title   = {Improved absolute frequency measurement of the strontium ion clock using a {GPS} link to the {SI} second},
  journal = {Metrologia},
  volume  = {60},
  number  = {1},
  pages   = {015007},
  year    = {2023},
  doi     = {10.1088/1681-7575/aca615}
}

@article{Sahoo2021,
  author  = {Sahoo, B.~K. and Das, B.~P. and Spiesberger, H.},
  title   = {New physics constraints from atomic parity violation in {$^{133}${Cs}}},
  journal = {Phys. Rev. D},
  volume  = {103},
  number  = {11},
  pages   = {L111303},
  year    = {2021},
  doi     = {10.1103/PhysRevD.103.L111303}
}

@article{Roberts2022,
  author  = {Roberts, B.~M. and Ginges, J.~S.~M.},
  title   = {Comment on {``New physics constraints from atomic parity violation in {$^{133}${Cs}}''}},
  journal = {Phys. Rev. D},
  volume  = {105},
  pages   = {018301},
  year    = {2022},
  doi     = {10.1103/PhysRevD.105.018301}
}

@article{Sahoo2022,
  author  = {Sahoo, B.~K. and Das, B.~P. and Spiesberger, H.},
  title   = {Reply to {``Comment on `New physics constraints from atomic parity violation in {$^{133}${Cs}}`''}},
  journal = {Phys. Rev. D},
  volume  = {105},
  pages   = {018302},
  year    = {2022},
  doi     = {10.1103/PhysRevD.105.018302}
}

@misc{Flambaum2026,
  author        = {Flambaum, V.~V. and Samsonov, I.~B.},
  title         = {Parity violation in atoms and electron scattering revisited},
  year          = {2026},
  eprint        = {2602.22466},
  archivePrefix = {arXiv},
  primaryClass  = {hep-ph}
}

@article{Katyal2025,
  author  = {Katyal, V. and Chakraborty, A. and Sahoo, B.~K. and Ohayon, B. and Seng, C.-Y. and Gorchtein, M. and Behr, J.},
  title   = {Testing for isospin symmetry breaking by combining isotope shift measurements with precise calculations in potassium},
  journal = {Phys. Rev. A},
  volume  = {111},
  pages   = {042813},
  year    = {2025},
  doi     = {10.1103/PhysRevA.111.042813}
}

@article{Kreim2014,
  author  = {Kreim, K. and Bissell, M.~L. and Papuga, J. and Blaum, K. and {De Rydt}, M. and {Garcia Ruiz}, R.~F. and Goriely, S. and Heylen, H. and Kowalska, M. and Neugart, R. and Neyens, G. and Nörtershäuser, W. and Rajabali, M.~M. and {Sánchez Alarcón}, R. and Stroke, H.~H. and Yordanov, D.~T.},
  title   = {Nuclear charge radii of potassium isotopes beyond N=28},
  journal = {Phys. Lett. B},
  volume  = {731},
  pages   = {97--102},
  year    = {2014},
  doi     = {10.1016/j.physletb.2014.02.012}
}

@article{konig_nuclear_2024,
  author  = {König, K. and Berengut, J.~C. and Borschevsky, A. and Brinson, A. and Brown, B.~A. and Dockery, A. and Elhatisari, S. and Eliav, E. and {Garcia Ruiz}, R.~F. and Holt, J.~D. and Hu, B.-S. and Karthein, J. and Lee, D. and Ma, Y.-Z. and Meißner, U.-G. and Minamisono, K. and Oleynichenko, A.~V. and Pineda, S.~V. and Prosnyak, S.~D. and Reitsma, M.~L. and Skripnikov, L.~V. and Vernon, A. and Zaitsevskii, A.},
  title   = {Nuclear Charge Radii of Silicon Isotopes},
  journal = {Phys. Rev. Lett.},
  volume  = {132},
  number  = {16},
  pages   = {162502},
  year    = {2024},
  doi     = {10.1103/PhysRevLett.132.162502}
}

@article{sanchez_nuclear_2006,
  author  = {Sánchez, R. and Nörtershäuser, W. and Ewald, G. and Albers, D. and Behr, J. and Bricault, P. and Bushaw, B.~A. and Dax, A. and Dilling, J. and Dombsky, M. and Drake, G.~W.~F. and Götte, S. and Kirchner, R. and Kluge, H.-J. and Kühl, T. and Lassen, J. and Levy, C.~D.~P. and Pearson, M.~R. and Prime, E.~J. and Ryjkov, V. and Wojtaszek, A. and Yan, Z.-C. and Zimmermann, C.},
  title   = {Nuclear Charge Radii of {$^{9,11}$Li}: The Influence of Halo Neutrons},
  journal = {Phys. Rev. Lett.},
  volume  = {96},
  number  = {3},
  pages   = {033002},
  year    = {2006},
  doi     = {10.1103/PhysRevLett.96.033002}
}

@article{sahoo_accurate_2010,
  author  = {Sahoo, B.~K.},
  title   = {Accurate estimate of {$\alpha$} variation and isotope shift parameters in {Na} and {Mg}$^{+}$},
  journal = {J. Phys. B: At. Mol. Opt. Phys.},
  volume  = {43},
  number  = {23},
  pages   = {231001},
  year    = {2010},
  doi     = {10.1088/0953-4075/43/23/231001}
}

@article{konig_performance_2020,
  author  = {König, K. and Imgram, P. and Krämer, J. and Maaß, B. and Mohr, K. and Ratajczyk, T. and Sommer, F. and Nörtershäuser, W.},
  title   = {On the performance of wavelength meters: Part 2—frequency-comb-based characterization for more accurate absolute wavelength determinations},
  journal = {Appl. Phys. B},
  volume  = {126},
  number  = {5},
  pages   = {86},
  year    = {2020},
  doi     = {10.1007/s00340-020-07433-4}
}

@book{King.1984,
	address = {New York, NY},
	series = {Physics of {Atoms} and {Molecules} {Ser}},
	title = {Isotope {Shifts} in {Atomic} {Spectra}},
	isbn = {978-1-4899-1786-7},
	url = {https://ebookcentral.proquest.com/lib/kxp/detail.action?docID=6709133},
	publisher = {Springer},
	author = {King, W. H.},
	month = jan,
	year = {1984},
}

@article{Manovitz2019,
  author  = {Manovitz, T.},
  title   = {Precision Measurement of Atomic Isotope Shifts Using a Two-Isotope Entangled State},
  journal = {Phys. Rev. Lett.},
  volume  = {123},
  number  = {20},
  pages   = {203001},
  year    = {2019},
  doi     = {10.1103/PhysRevLett.123.203001}
}

@book{Moore1971,
  author    = {Moore, C.~E.},
  title     = {Atomic Energy Levels as Derived from the Analyses of Optical Spectra -- Chromium through Niobium},
  series     = {Nat. Stand. Ref. Data Ser., NSRDS-NBS},
  volume     = {35},
  number     = {2},
  publisher  = {National Bureau of Standards, U.S.},
  address    = {Washington, D.C.},
  year       = {1971},
  note       = {Reprint of NBS Circ. 467, Vol. II (1952)},
  pages      = {230},
  doi        = {10.6028/NBS.NSRDS.35v2}
}

@article{Cheal2012,
  author = {Cheal, B. and Flanagan, K. T.},
  title = {Laser spectroscopy of radioactive isotopes: Role and limitations of accurate isotope-shift calculations},
  journal = {Phys. Rev. A},
  volume = {86},
  number = {4},
  pages = {042501},
  year = {2012},
  doi = {10.1103/PhysRevA.86.042501},
}

@article{Filippin2016,
  author = {Filippin, Livio and Beerwerth, Randolf and Ekman, J{\"o}rgen and Fritzsche, Stephan and Godefroid, Michel and J{\"o}nsson, Per},
  title = {Multiconfiguration calculations of electronic isotope shift factors in {Al} {I}},
  journal = {Phys. Rev. A},
  volume = {94},
  number = {6},
  pages = {062508},
  year = {2016},
  month = dec,
  doi = {10.1103/PhysRevA.94.062508},
}

@article{Ohayon2022,
  author = {Ohayon, B. and Garcia Ruiz, R. F. and Sun, Z. H. and Hagen, G. and Papenbrock, T. and Sahoo, B. K.},
  title = {Nuclear charge radii of {Na} isotopes: Interplay of atomic and nuclear theory},
  journal = {Phys. Rev. C},
  volume = {105},
  number = {3},
  pages = {L031305},
  year = {2022},
  month = mar,
  doi = {10.1103/PhysRevC.105.L031305},
}

@article{Berengut.2025,
	title = {Second-order hyperfine structure and its impact on searches for new physics using isotope-shift spectroscopy},
	volume = {112},
	doi = {10.1103/qkv8-z8z1},
	number = {2},
	journal = {Physical Review A},
	author = {Berengut, Julian C.},
	year = {2025},
}

@misc{Angeli.2026,
      title={Towards better nuclear charge radii}, 
      author={István Angeli and Dimiter L. Balabanski and Paraskevi Dimitriou and Dipti and Kieran T. Flanagan and Georgi Georgiev and Mikhail Gorchtein and Paul Gùeye and Fabian Heiße and Andreas Knecht and Kei Minamisono and Wilfried Nörtershäuser and Ben Ohayon and Natalia S. Oreshkina and B. K. Sahoo and Hunter Staiger and Endre Takacs and Xiaofei Yang and Deyan T. Yordanov},
      year={2026},
      eprint={2604.08985},
      archivePrefix={arXiv},
      doi = {10.48550/arxiv.2604.08985}
}

@misc{Beyer2025,
  author        = {Konstantin A. Beyer and Igor A. Valuev and Zoia A. Mandrykina and Zewen Sun and Natalia S. Oreshkina},
  title         = {Relativistic Recoil as a Key to the Fine-Structure Puzzle in Muonic $^{90}${Zr}},
  year          = {2025},
  eprint        = {2511.22298},
  archivePrefix = {arXiv},
  primaryClass  = {physics.atom-ph},
  note          = {arXiv:2511.22298}
}

@misc{Rathi2026,
  author        = {S. Rathi and I. A. Valuev and Z. Sun and M. Heines and P. Indelicato and B. Ohayon and N. S. Oreshkina},
  title         = {Theory Framework for Medium-Mass Muonic Atoms},
  year          = {2026},
  eprint        = {2603.22021},
  archivePrefix = {arXiv},
  primaryClass  = {physics.atom-ph},
  note          = {arXiv:2603.22021}
}

@article{Krieger2017,
  author  = {Krieger, A. and N{\"o}rtersh{\"a}user, W. and Geppert, Ch. and Blaum, K. and Bissell, M. L. and Fr{\"o}mmgen, N. and Hammen, M. and Kreim, K. and Kowalska, M. and Kr{\"a}mer, J. and Neugart, R. and Neyens, G. and S{\'a}nchez, R. and Tiedemann, D. and Yordanov, D. T. and Zakova, M.},
  title   = {Frequency-comb referenced collinear laser spectroscopy of {Be}$^{+}$ for nuclear structure investigations and many-body {QED} tests},
  journal = {Applied Physics B},
  volume  = {123},
  number  = {1},
  pages   = {15},
  year    = {2017},
  doi     = {10.1007/s00340-016-6579-5}
}

\end{document}